\documentclass[fleqn,usenatbib]{mnras}

\usepackage{newtxtext,newtxmath}

\usepackage[T1]{fontenc}
\usepackage{ae,aecompl}
\usepackage[utf8]{inputenc}

\usepackage{graphicx}	
\usepackage{amsmath}	
\usepackage{amssymb}	
\usepackage{siunitx}    
\usepackage{booktabs,caption}
\usepackage[flushleft]{threeparttable}
\usepackage{enumerate}

\title[Testing the blast-wave AGN feedback scenario]{Testing the blast-wave AGN feedback scenario in MCG-03-58-007}

\author[Mattia Sirressi et al.]{
M. Sirressi$^{1,2}$\thanks{E-mail: m.sirressi@campus.unimib.it},
C. Cicone$^{2}$,
P. Severgnini$^{2}$,
V. Braito$^{3,4}$,
M. Dotti$^{1}$,
R. Della Ceca$^{2}$,
\newauthor{
J.N. Reeves$^{4}$,
G.A. Matzeu$^{3,5}$,
C. Vignali$^{6,7}$,
L. Ballo$^{5}$
}
\\
$^{1}$Università degli Studi di Milano-Bicocca, Dip. di Fisica G. Occhialini, Piazza della Scienza 3, 20126 Milano, Italy\\
$^{2}$INAF - Osservatorio Astronomico di Brera, Via Brera 28, I-20121, Milano, Italy\\
$^{3}$INAF – Osservatorio Astronomico di Brera, Via Bianchi 46, I-23807 Merate (LC), Italy\\
$^{4}$Center for Space Science and Technology, University of Maryland Baltimore County, 1000 Hilltop Circle, Baltimore, MD 21250, USA\\
$^{5}$European Space Astronomy Centre (ESA/ESAC), E-28691 Villanueva de la Canada, Madrid, Spain\\
$^{6}$Dipartimento di Fisica e Astronomia, Università di Bologna, via Gobetti 93/2, I–40129 Bologna, Italy\\
$^{7}$INAF - Osservatorio di Astrofisica e Scienza dello Spazio di Bologna, via Gobetti 93/3, I–40129 Bologna, Italy\\}

\date{Accepted XXX. Received YYY; in original form ZZZ}

\pubyear{2019}

\begin{document}
\label{firstpage}
\pagerange{\pageref{firstpage}--\pageref{lastpage}}
\maketitle

\begin{abstract}

We report the first Atacama large millimeter/submillimeter array observations of MCG-03-58-007, a local ($z=0.03236\pm0.00002$, this work) AGN ($L_{AGN}\sim10^{45}~\rm erg~s^{-1}$), hosting a powerful X-ray ultra-fast ($v=0.1c$) outflow (UFO). The CO(1-0) line emission is observed across $\sim18$~kpc scales with a resolution of $\sim 1\,\rm kpc$. About 78\% of the CO(1-0) luminosity traces a galaxy-size rotating disk. However, after subtracting the emission due to such rotating disk, we detect with a S/N=20 a residual emission in the central $\sim 4$~kpc. Such residuals may trace a low velocity ($v_{LOS}=170$~km~s$^{-1}$) outflow. We compare the momentum rate ($\dot{P}$) and kinetic power ($\dot{E}$) of such putative molecular outflow with that of the X-ray UFO and find $\dot{P}_{mol}/\dot{P}_{UFO}\sim0.4$ and $\dot{E}_{mol}/\dot{E}_{UFO}\sim4\cdot10^{-3}$. This result is at odds with the energy-conserving scenario suggested by the large momentum boosts measured in some other molecular outflows. An alternative interpretation of the residual CO emission would be a compact rotating structure, distinct from the main disk, which would be a factor of $\sim10-100$ more extended and massive than typical circumnuclear disks revealed in Seyferts. In conclusion, in both scenarios, our results rule out the hypothesis of a momentum-boosted molecular outflow in this AGN, despite the presence of a powerful X-ray UFO.

\end{abstract}

\begin{keywords}
galaxies: evolution -- galaxies: active -- galaxies: individual (MCG-03-58-007)
\end{keywords}



\section{Introduction}\label{sec:intro}



In the context of galaxy formation and evolution studies, it is crucial to understand the physics of the feedback processes from Active Galactic Nucleus (AGN) and Star Formation (SF) activity, which can manifest through powerful galactic outflows \citep{Silk1998A&A...331L...1S,Sijacki2007MNRAS.380..877S}. AGN feedback processes may have produced the $\rm M_{BH}-\sigma_{*}$ relation \citep{2013ARA&A..51..511K} through a mechanical interaction between the AGN and the host galaxy. Such an interaction can be provided by powerful and fast winds that develop from the accretion disk and can affect the properties of the interstellar medium (ISM), transporting energy and momentum up to the kpc scales \citep[blast-wave AGN feedback models, e.g.][]{Silk1998A&A...331L...1S,king2010black}. Furthermore, feedback mechanisms, from both AGN and SF activity, are often invoked to justify additional observational properties of galaxies: the shape of the baryonic mass function of galaxies \citep{Papastergis2012ApJ...759..138P}, the SF quenching of massive galaxies as indicated by the galaxy colour bimodality \citep{DiMatteo2005Natur.433..604D,Schawinski2014MNRAS.440..889S}, and the $[\alpha/Fe]$-enhancement in massive spheroids \citep{Fontanot2009MNRAS.397.1776F}.

X-ray high signal-to-noise observations led to the discovery of ultra-fast ($v>0.1c$) outflows (UFO) in 40\% of the bright nearby local AGN population \citep{Gofford2015MNRAS.451.4169G}. The X-ray spectra of these sources show one or more deep absorption lines that are best identified with the transitions of highly ionised Iron \citep{Reeves2003ApJ...593L..65R,Tombesi2010ApJ...719..700T,Gofford2015MNRAS.451.4169G}. These transitions, modelled with photoionisation models, are blue-shifted with velocities from a few thousand $\rm km\,s^{-1}$ up to tens of thousands of $\rm km\,s^{-1}$ and therefore provide strong evidence for the presence of fast outflowing absorbers at sub-pc scales. Often classified as disk-winds, due to their connection with the accretion disk, X-ray UFOs are a direct manifestation of AGN feedback mechanisms in action.

Another revolution in the study of feedback processes in galaxies has been the recent discovery of massive, galaxy-size, cold molecular outflows with velocities in the range between a few hundreds and a few thousands $\rm km\,s^{-1}$  \citep{Feruglio2010A&A...518L.155F,Fischer2010A&A...518L..41F}. Less massive, multi-phase galactic winds were known since the 1980's including ionised, neutral and molecular gas components \citep{Nakai1987PASJ...39..685N}. However, the massive molecular outflows discovered in the recent decade are estimated to embed up to $\rm M_{mol}\sim10^{10}$~M$_{\odot}$ of $\rm H_2$ gas  \citep{Cicone2018ApJ...863..143C}. Their signature has been found in far infrared (FIR) spectra featuring prominent OH and H$_2$O P-Cygni profiles \citep{Sturm2011ApJ...733L..16S,Veilleux2013ApJ...776...27V,Spoon2013ApJ...775..127S,Gonzalez-Alfonso+17} as well as in mm-interferometric observations of molecular emission lines (CO, HCN, HCO$^+$) showing high velocity blue- and red-shifted components clearly deviating from the disk rotation (e.g. \citealt{Alatalo+11,Aalto+12,Cicone2012A&A...543A..99C, Combes+13, Cicone2014A&A...562A..21C, Garcia-Burillo+14}).

Among the theoretical models of AGN-driven outflows that have been proposed, the blast-wave feedback scenario aims to understand the dynamics of both nuclear disk-winds and galactic-scale outflows \citep{Silk1998A&A...331L...1S,king2010black,Faucher2012MNRAS.425..605F}. According to this class of models, a nuclear wind interacts with the ISM, producing a reverse shocked wind and at the same time sweeping up the shocked ambient medium, thus originating a large-scale outflow. One open question regarding this model is whether the outflow is energy-driven (i.e. energy-conserving) or momentum-driven. 
From a theoretical perspective, there is not a well established agreement on which outflow regime is favoured in nature. If the cooling of the shocked nuclear wind is efficient, the blast-wave models predict a momentum-driven large-scale outflow whose momentum rate is by definition the same as that of the nuclear wind, i.e.  $\dot{P}_{out}\simeq \dot{P}_{wind}\simeq L_{AGN}/c$ \citep{king2010black}. If the Compton cooling of the shock is instead inefficient, the energy-driven large-scale outflow should show a momentum boost of $\dot{P}_{out}\gtrsim10\cdot\dot{P}_{wind}$ \citep{Faucher2012MNRAS.425..605F}. Some of the galaxy-scale molecular outflows detected so far in ultra luminous infrared galaxies (ULIRGs) have a momentum rate that exceeds by an average factor of 20 the momentum flux of the AGN photons \citep{Cicone2014A&A...562A..21C,Fiore2017A&A...601A.143F}, which would favour a blast-wave energy-driven scenario. For example, in Mrk~231 \citep{Feruglio2015A&A...583A..99F} and IRAS~F11119+3257 \citep{Tombesi2015Natur.519..436T}, the momentum rate of the molecular outflow appears to exceed that of the X-ray wind by a factor of several. However, such measurements are highly uncertain. Indeed, in IRAS~F11119+3257 the molecular outflow energetics revisited using new ALMA data appears more consistent with the momentum conserving scenario \citep{Veilleux2017ApJ...843...18V}. Another notable recent counter example is the quasar PDS~456, where energy-conserving feedback seems to have been ruled out \citep{Bischetti2019arXiv190310528B}. Increased sample sizes are delivering a much broader range of momentum-boost values \citep{Garcia-Burillo+15,Pereira-Santaella+18,Fleutsch2019MNRAS.483.4586F}.

Another mechanism that can drive galactic outflows, without necessarily requiring the presence of a fast nuclear wind, is based on the AGN radiation pressure acting on the dust that is dynamically coupled at least to the cold phase of the ISM \citep{Costa2018MNRAS.479.2079C}. In the multi-scattering regime, which requires the dust to be optically thick at IR wavelengths, the momentum rate of the large-scale outflow triggered with such mechanism can exceed the photon momentum \citep{IshFab2015MNRAS.451...93I,Bieri2017MNRAS.464.1854B,Costa2018MNRAS.479.2079C}. Both analytically \citep{Ishibashi2018MNRAS.476..512I} and through hydrodynamical simulations \citep{Costa2018MNRAS.479.2079C}, it has been shown that AGN radiative feedback models in luminous quasars ($L_{AGN}\sim10^{45-47}\,\rm erg\,s^{-1}$) can explain the observed molecular outflows with a moderate momentum boost. However, the ability of radiation pressure to significantly accelerate gas-shells is limited to small ($<1\,\rm kpc$) scales where the gas is optically thick in the IR \citep{Tiago2018MNRAS.473.4197C}.
When compared to thermal feedback models, IR radiation pressure can more efficiently eject gas from the bulge and hence regulate SF, but only as long as the gas density is high enough: when a large fraction of gas is blown away, the ISM becomes optically thin and the radiation pressure inefficient \citep{Costa2018MNRAS.479.2079C}.

So far, most molecular outflow studies have focused on ULIRGs, whose high infrared luminosity is produced by a combination of AGN and SF activity. In such extreme objects, the intense SF activity, with $SFR \gtrsim 100\,\rm M_{\odot}\,yr^{-1}$, is responsible for an important contribution to the feedback processes that is difficult to disentangle from the AGN feedback processes \citep{Cicone2014A&A...562A..21C}. To partially overcome this limitation, in this work we study MCG-03-58-007, which is a local (z=0.03) Seyfert 2, classified as LIRG. The source has a moderate SFR of $20\, \rm M_{\odot}\,yr^{-1}$ \citep{Gruppioni201610.1093/mnras/stw577}, and at the same time hosts a powerful X-ray UFO \citep{braito2018new}, hence we expect AGN feedback processes to dominate over SF feedback. Indeed, the kinetic power of the X-ray wind ($\dot{E}_k=8\%\,\rm L_{AGN}$) exceeds the theoretical threshold required to have feedback from an AGN \citep{Hopkins2010MNRAS.401....7H}. Such properties make MCG-03-58-007 one of the best candidates for isolating the AGN feedback from the SF contribution with the aim of testing the blast-wave feedback models.

The paper is organised as follows. In Section \ref{sec:target} we describe the target galaxy and summarise the properties of the X-ray UFO. In  Section~\ref{sec:ALMA} we present the ALMA observations focusing on the CO(1-0) line. The analysis of the data is presented in Section \ref{sec:data analysis}, where we describe the rotating disk model constructed with the software \textsc{3D Barolo}. In Section \ref{sec:discussion} we discuss the interpretation of the results and test the AGN feedback models. In Section \ref{sec:conclusions} we summarise our findings and their implications for the current understanding of the feedback mechanisms at work in galaxies.

Throughout this work, a $H_0 = 70$ \si{km.s^{-1}.Mpc^{-1}}, $\Omega_{\Lambda_0} = 0.73$ and $\Omega_m = 0.27$ Cosmology is adopted. At the source redshift ($z=0.03236$) $1\arcsec$ corresponds to 0.67 kpc. Errors are given at 1$\sigma$ confidence level unless otherwise specified, i.e. Section \ref{sec:disk} and Table \ref{tab:residuals}.

\section*{}
\addtocounter{section}{1}






\subsection{The target galaxy: MCG-03-58-007}\label{sec:target}
MCG-03-58-007 is a nearby ($z = 0.03236$, redshift measurement refined in this work, see section \ref{sec:data reduction}) Seyfert type 2 galaxy ($L_{AGN}\sim3\cdot10^{45}\rm erg\,s^{-1}$(\footnote{Throughout this work, $L_{AGN}$ is the bolometric luminosity of the AGN}), for details see \citealt[]{braito2018new}) and LIRG ($L_{IR}=2.3\cdot10^{11}L_{\odot}$, \citealt{Gruppioni201610.1093/mnras/stw577}). 
A star formation rate of  SFR=$20.1\pm0.9\,\rm M_{\odot}~yr^{-1}$ was estimated by \citet{Gruppioni201610.1093/mnras/stw577}, by using the \citet{Kennicutt1998ApJ...498..541K} relation converted to a Chabrier IMF.

Direct evidence for AGN feedback in this source was found with the discovery of an X-ray highly ionized UFO with $v \gtrsim 0.1c$ and log$(\xi/\rm erg\,cm\,s^{-1})\,\sim5.5$ (\footnote{The ionisation parameter is defined as $\xi = \frac{L_{ion}}{n_{\rm e}R^2}$, where $L_{ion}$
is the ionising luminosity in the 1-1000 Rydberg range, $R$ is the
distance to the ionising source and $n_{\rm e}$ is the electron density. The
units of $\xi$ are $\rm erg\,cm\,s^{-1}$. $1\,Ry=13.6\,\rm eV$.}) in the Suzaku spectrum in 2010 \citep{braito2018new,Matzeu2019MNRAS.483.2836M}. 
Two deep absorption troughs at $E=7.4\,\rm keV$ and $E=8.5\,\rm keV$, which can be associated with the blue-shifted transitions of Fe XXV and Fe XXVI, strongly indicate the presence of two highly ionized, high column density and fast outflowing absorbers with velocities of $v_1\sim0.1c$ and $v_2\sim0.2c$. Follow-up XMM-Newton and NuSTAR observations in 2015 confirmed the presence of the $v_1=0.1c$ zone of the wind and possibly of an even faster zone \citep{braito2018new}. These  observations also showed an occultation event consistent with an increase of the disk-wind opacity. 

\cite{braito2018new} compared the energetics of the nuclear wind to the bolometric luminosity of the AGN. They considered only the wind component moving at $v_1 \sim 0.1c$, being the only one persistent over time. The mass of the central black hole expected from the $M_{BH}-\sigma_{*}$ relation is $M_{BH}\sim 10^8\, \rm M_{\odot}$. The mass-loss rate of the X-ray wind ($\dot{M}\sim1.5\,\rm M_{\odot}\,yr^{-1}$) was calculated assuming a biconical geometry for the flow. The kinetic power of the persistent component of the wind is $\dot{E}_{k1} \sim 2.4\cdot10^{44}\,\rm erg\,s^{-1}$ ($8\%\,L_{AGN}$) and the momentum rate is $\dot{p}_1 \sim 2\cdot 10^{35}\,\rm g\,cm\,s^{-2}$, which is of the same order of $L_{AGN}/c$. Since the covering factor and the launching radius of the wind are difficult to determine, the uncertainties of the momentum rate and kinetic power are at least $40-50\%$. 

Besides hosting a stable X-ray UFO, MCG-03-58-007 also hosts an ionised outflow with $v_{ion}\sim500\,\rm km\,s^{-1}$, indicated by the broad blue-shifted wings of the [OIII]4959$\lambda$ and [OIII]5007$\lambda$ lines visible in the optical spectrum as mentioned by \citet{braito2018new}.

\subsection{ALMA observations}\label{sec:ALMA}
\begin{figure}
    \centering
    \includegraphics[width=\columnwidth]{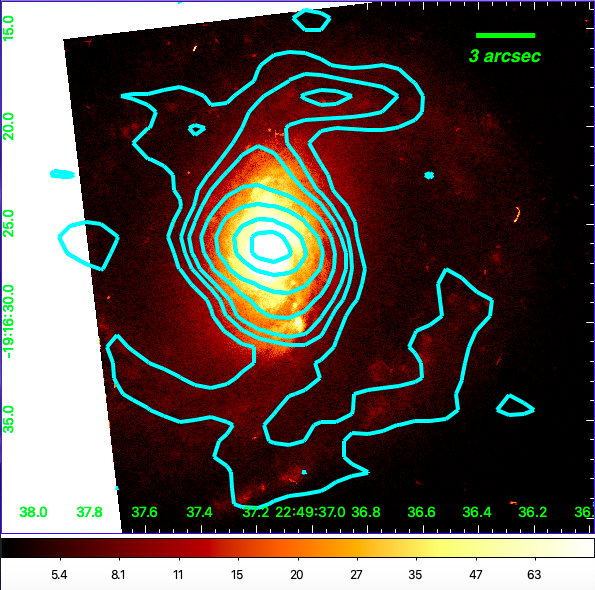}
    \caption{ALMA CO(1-0) emission map (cyan contours) superimposed to the Hubble Space Telescope (HST) optical image (F606W filter) of MCG-03-58-007. The horizontal and vertical axes display the Right Ascension (RA) and Declination (Dec) respectively, in sexagesimal units. The CO(1-0) contours levels correspond to 1-3-5-10-20-40-60 $\sigma$, with $\sigma=0.6\,\rm mJy$. The optical image is scaled with the inverse hyperbolic sine function (asinh). The extension of the galaxy optical emission coincides with the CO(1-0) line emission and is about $25\arcsec$ in diameter. The synthesized beam size of the ALMA data is $3.3\arcsec x 2.0\arcsec$.}
    \label{fig:hst-CO}
\end{figure}

We observed MCG-03-58-007 with ALMA Band 3 during Cycle 4 (Project code: 2016.1.00694.S, PI: P. Severgnini). We used a compact configuration of 40 12-m antennas, in a single pointing with field of view of $75\arcsec$ (FWHM), which fully covered the target galaxy, as shown in Figure \ref{fig:hst-CO}.
The observations were carried out on 11 March 2017 with an on-source time of $\rm 7.4\,min$. 

We arranged a spectral setup consisting of 2 spectral windows in each sideband. In the upper sideband we placed one spectral window centred at $\nu_{obs}=111.682\,\rm GHz$ for detecting the CO~J=1-0 (hereafter CO(1-0)) line, and another spectral window centred at $\nu_{obs}=109.952\,\rm GHz$ to sample the CN (N=1-0, J=3/2-1/2, J=1/2-1/2) line doublet, which will be analysed in a separate publication. Both spectral windows in the upper sideband have a width of 1.875 GHz and a resolution of 1.95 MHz. In the lower sideband we placed two spectral windows of 2 GHz width and 15.6 MHz resolution centred at $\nu_{obs}=97.952\,\rm GHz$ and $\nu_{obs}=99.807\,\rm GHz$, to study the 3-mm continuum emission. Table \ref{tab:ALMA-data log} lists the main parameters of the ALMA CO(1-0) observations. 
Three calibrators (J2258-2758, Uranus, J2236-1433) were used for fixing the flux density scale, for determining the bandpass response, and for calibrating amplitude and phase of the visibilities of the science target. The calibration pipeline was launched with \textsc{CASA} version 4.7, to produce the visibility files. The data reduction and analysis were performed with \textsc{CASA} version 5.4. 

The CO(1-0) line emission image was deconvolved using the \texttt{TCLEAN} task with Briggs weighting (robust = 0.0). We investigated different masking procedures: in order to create a different mask for every channel, we ultimately used the \texttt{AUTOMASK} parameter. The latter automatically provides a different mask for each channel and minimizes the residual negative sidelobes. The cleaning procedure was carried out over the spectral range corresponding to velocities of $v\in(-1000, +1000)$~km~s$^{-1}$, on the basis of a preliminary quick imaging that shows no line emission at higher velocities (see also Section~\ref{sec:data analysis}). We adopted the native spectral resolution of 5 $\rm km\,s^{-1}$ wide channels. Frequencies were converted to velocities using the optical convention and using the corrected redshift value measured in this work (see Section \ref{sec:data reduction}). We chose a cleaning threshold equal to the rms per channel (2.5~mJy) and a pixel size of 0.2$\arcsec$. The primary beam correction was included in the cleaning task to account for the dependence of the sensitivity on the direction within the field of view.
The final synthesized beam has an average size of $3.3\arcsec$ x $2.0\arcsec$ which corresponds to a physical spatial resolution of $2.2\,$kpc x $1.3\,$kpc. The achieved rms sensitivity is $2.2\,\rm mJy / beam / 5.2\,\rm km\,s^{-1}$, as measured from channels free of line emission.

\begin{table}
    \centering
	\caption{ALMA C0(1-0) observation log}
	\label{tab:ALMA-data log}
	\begin{tabular}{cc} 
		\hline
		\multicolumn{2}{c}{Target line CO(1-0)}\\
		\hline
		$\nu_{obs}$ & 111.675 GHz\\
		Date & 11 March 2017\\
		Array configuration & C40-1\\
		On-source time & $\rm 6.70\,min$\\
		Channel width & 5.2 $\rm km\,s^{-1}$\\
		Spectral window's width & 4992 $\rm km\,s^{-1}$\\
		Rms sensitivity & $2.2\,\rm mJy / beam / 5.2\,\rm km\,s^{-1}$ \\
		Synthesized beam &  $3.3\arcsec x 2.0\arcsec$\\
		\hline
	\end{tabular}
\end{table}

\section{Analysis and results}\label{sec:data analysis}

\subsection{Continuum subtraction and redshift correction}\label{sec:data reduction}

\begin{figure}
	\includegraphics[width=\columnwidth]{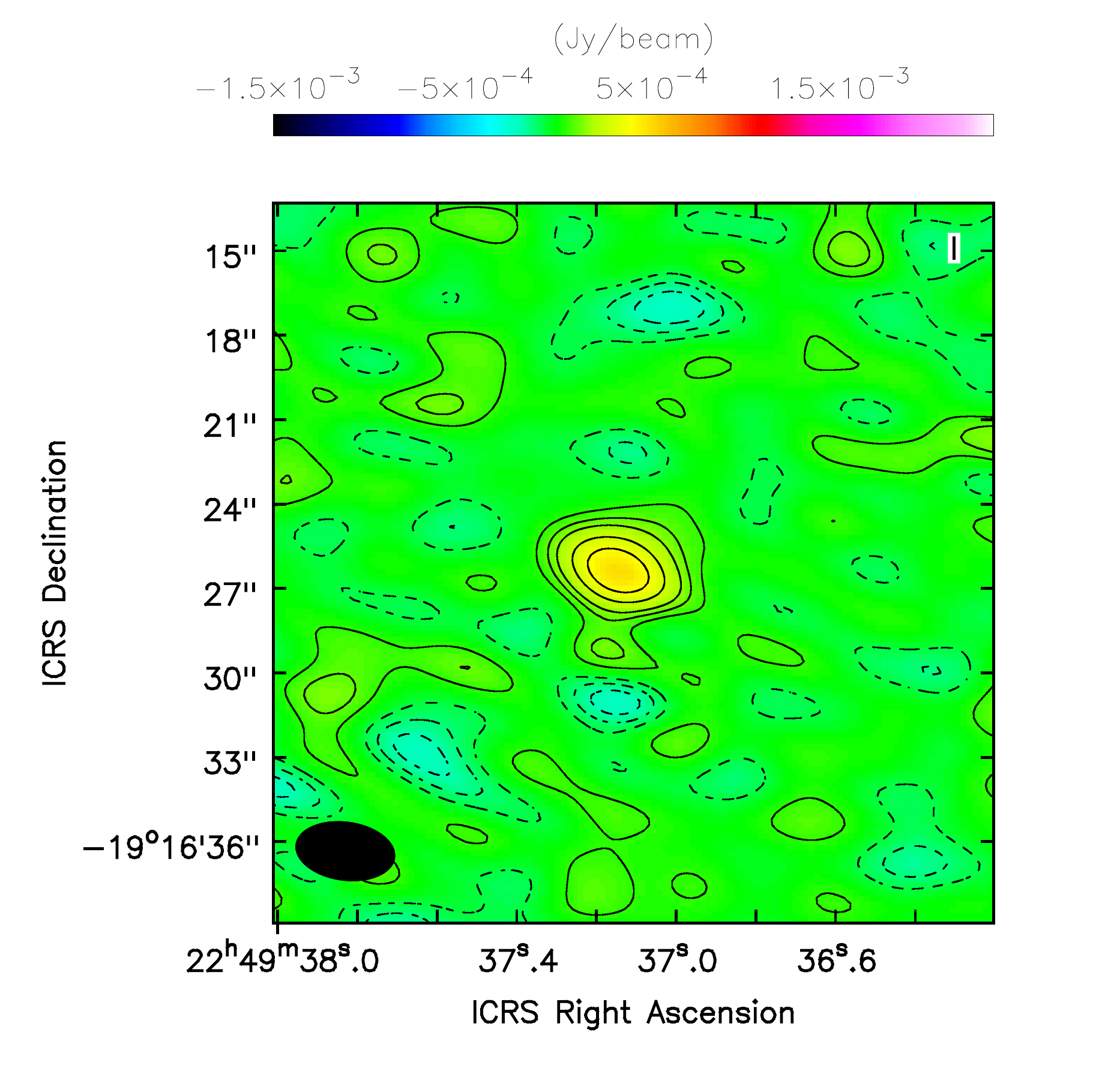}
    \caption{Continuum emission of MCG-03-58-007 at an average $\nu_{obs}=99\,\rm GHz$. Two spectral windows, placed on the central frequencies $\nu_{obs}=98\,\rm GHz$ and $\nu_{obs}=100\,\rm GHz$, free of lines emission, were merged and cleaned for producing this map. The emission is centred on the coordinates of the target, the synthesized beam is represented by the black ellipse in the bottom-left corner of the figure. The black contours levels correspond to 1,2,3,5,7 $\sigma_{rms}$ with $\sigma_{rms}=0.065\,\rm  mJy$. Symmetric negative contours are plotted with dashed lines.}
    \label{fig:contMap}
\end{figure}

The 3-mm continuum emission was subtracted using the task \texttt{UVCONTSUB} from the {\it uv} visibilities. For the purpose of continuum subtraction, we estimated the continuum level  from the channels adjacent to the CO(1-0) line corresponding to velocities (-2500, -500) $\rm km\,s^{-1}$ and (500, 2500) $\rm km\,s^{-1}$. However, the 3-mm continuum emission is overall faint compared to the line emission. We measure a continuum flux density of $0.60\pm0.10$~mJy at $\nu_{obs}=99\,\rm GHz$. A map is shown in Figure \ref{fig:contMap}.

\begin{figure}
	\includegraphics[width=\columnwidth]{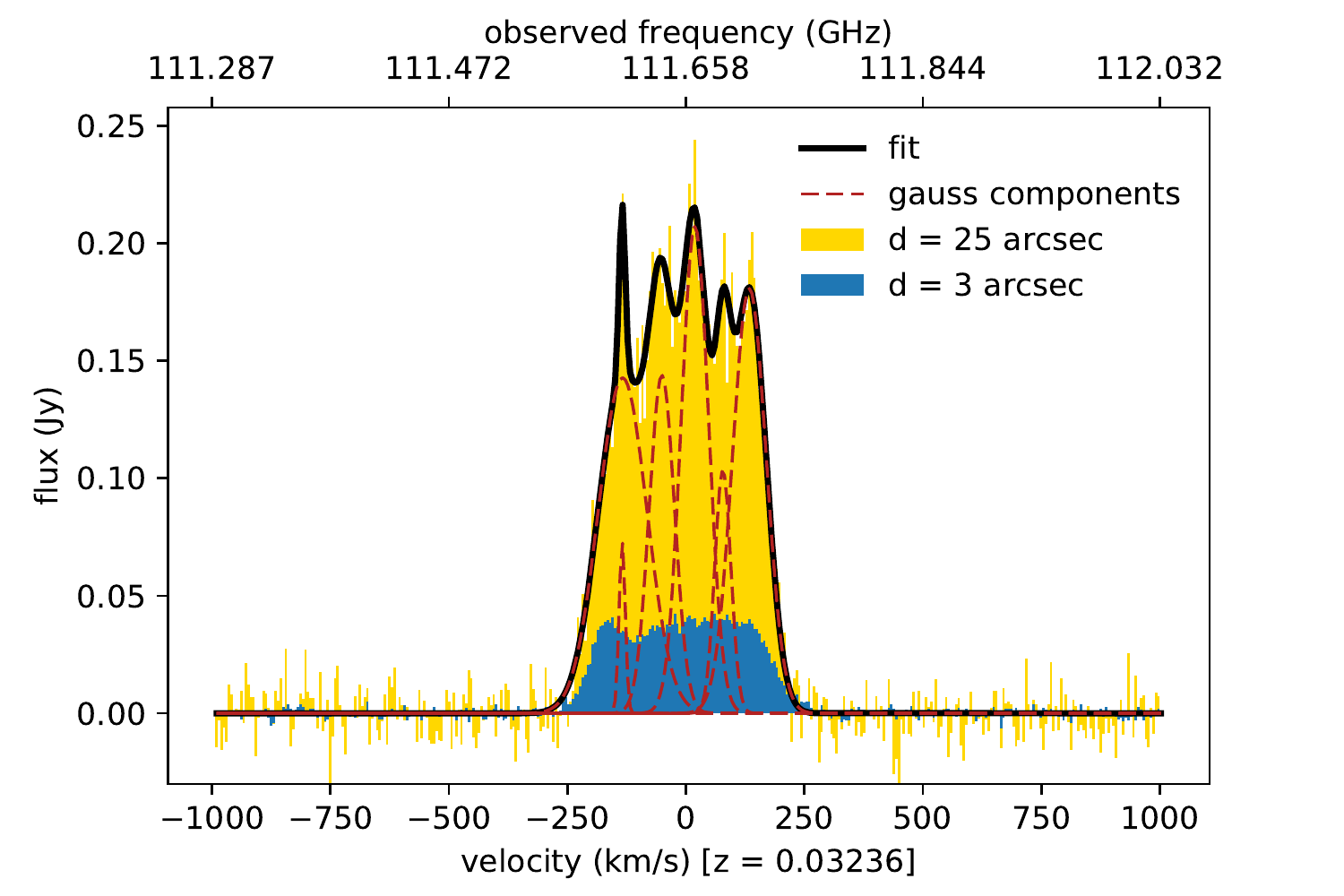}
    \caption{Continuum-subtracted spectrum of the CO(1-0) line in MCG-03-58-007,  extracted from circular apertures of $3\arcsec$ and of $25\arcsec$ (blue and yellow areas, respectively). The smaller aperture corresponds to the size of synthesized beam, the larger aperture covers the full galaxy extension and is the aperture that maximises the CO flux. The black solid line is the best fit of the total spectrum obtained with multiple Gaussian components (red dashed lines). The observed frequency (upper x-axis) has been converted to velocity (lower x-axis) using the redshift value refined in this work, $z=0.03236\pm0.00002$.}
    \label{fig:spectrum}
\end{figure}

Figure \ref{fig:spectrum} shows the continuum-subtracted spectrum of the CO(1-0) line, which is centred at the observed frequency $\nu_{obs} = 111.658\,\rm GHz$. This frequency is slightly different from that reported in Table \ref{tab:ALMA-data log} ($\nu_{obs} = 111.675\,\rm GHz$), which was our input CO(1-0) central frequency inferred from the CO(1-0) and CO(2-1) observations carried out with the Swedish European Southern Observatory (ESO) Submillimetre Telescope (SEST) and reported by \citet{Strong2004MNRAS.353.1151S}. These older millimetre data have a spectral resolution of $\Delta v = 30\,\rm km\,s^{-1}$, which is a factor of six larger than that of our ALMA data. 

\begin{figure}
	\includegraphics[width=\columnwidth]{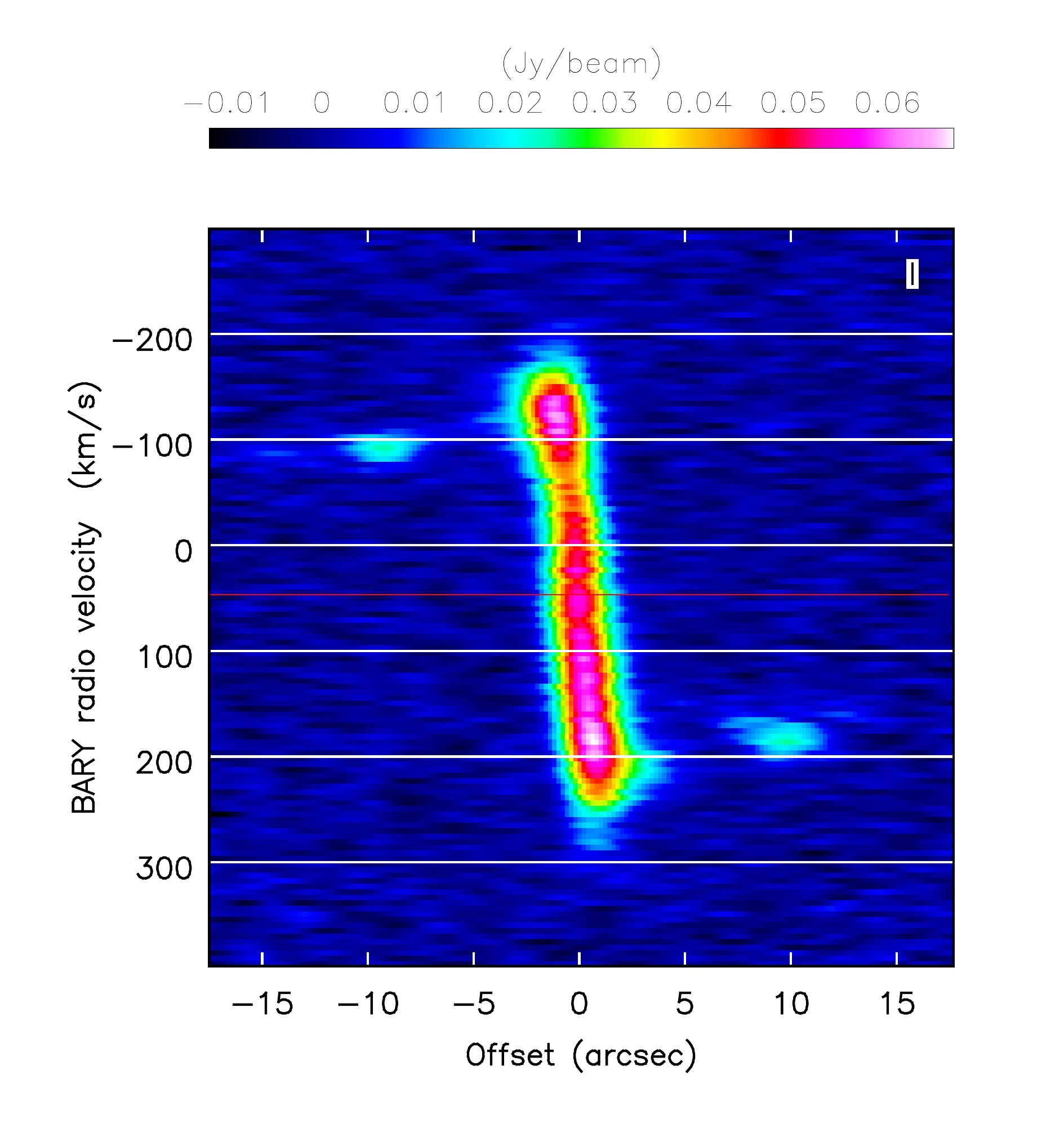}
    \caption{Position-velocity diagram extracted from a beam-wide slit with PA = 42$^\circ$. The red horizontal line marks the centroid velocity of the rotation curve, which is offset by $47\pm7\,\rm km\,s^{-1}$ with respect to the input redshift of $z=0.03220$. Our new redshift estimate is therefore $z=0.03236\pm0.00002$.}
    \label{fig:PV}
\end{figure}

In order to refine the redshift measurement, we used the CO(1-0) position velocity diagram extracted along an axis with a Position Angle (PA) of 42$^\circ$ (see Section \ref{sec:disk}), shown in Figure \ref{fig:PV}. From this plot we identified the centroid velocity of the rotation curve, which is offset by $v=47\pm7$ $\rm km\,s^{-1}$ with respect to the older redshift of $z=0.03220$ measured by \citet{Strong2004MNRAS.353.1151S}. By setting the centroid of the CO(1-0) rotation curve equal to the systemic velocity of the galaxy, we obtain a new redshift estimate of $z=0.03236\pm0.00002$ for MCG-03-58-007.

\subsection{Morphology of the CO line emission}\label{sec:morphology}

The CO(1-0) line spectrum extracted from a circular aperture of $25\arcsec$ shows a complex profile characterised by multiple components (Figure \ref{fig:spectrum}). For this reason we modelled the spectrum with multiple Gaussian functions to measure the total flux and luminosity of the CO(1-0) line. In Table \ref{tab:tot_flux} we report the best-fit parameters for each of the Gaussian components and the values of the CO(1-0) total flux, luminosity and total molecular gas mass. We derived a molecular gas mass of $M_{H2}\simeq(9 \pm 6)\frac{\alpha_{CO}}{3.1} \cdot 10^9\,\rm M_{\odot}$ using the formula given by \citet{solomon2005molecular} with a CO-to-$\rm H_2$ conversion factor of $\alpha_{CO}=3.1\pm2.1\rm\, M_{\odot}(K\,km\,s^{-1}\,pc^{2})^{-1}$. Such conversion factor is the average of the values calculated in nearby star-forming galaxies by \citet{Sandstrom2013ApJ...777....5S} and its uncertainty corresponds to the standard deviation of the sample.

\begin{table}
    \begin{threeparttable}
    \centering
	\caption{Results of a multi-Gaussian fit to the total CO(1-0) spectrum}
	\label{tab:tot_flux}
	\begin{tabular}{lccc} 
	    \hline
		Gaussian & $\mu_v$ & $S_{\rm peak}$ & $\sigma_v$ \\
		comp.   &   [$\rm km\,s^{-1}$] & [Jy] & [$\rm km\,s^{-1}$] \\
		\hline
		1 & $-133 \pm 7$ & $0.143 \pm 0.008$ & $73 \pm 6$\\
		2 & $-134 \pm 1$ & $0.073 \pm 0.010$ & $9.2 \pm 1.6$\\
		3 & $-50 \pm 5$ & $0.14 \pm 0.03$ & $38 \pm 8$\\
		4 & $19 \pm 3$ & $0.207 \pm 0.008$ & $41 \pm 8$\\
		5 & $78 \pm 3$ & $0.10 \pm 0.03$ & $24 \pm 4$\\
		6 & $134 \pm 3$ & $0.181 \pm 0.005$ & $50 \pm 3$\\
		\hline
		Total  & $S_{\rm CO}$ & $L^{\prime}_{\rm CO}$ & M$_{H2}$ \\
		spectrum & [Jy $\rm km\,s^{-1}$] & [$\rm{K\,km\,s^{-1}\,pc^{2}}$] & [$M_{\odot}$] \\
		\hline
		 & 65 $\pm$ 4 & (2.9 $\pm$ 0.5) $\cdot 10^9$  & (9 $\pm$ 6)$\frac{\alpha_{CO}}{3.1} \cdot 10^9$ \\
		\hline
	\end{tabular}
	\begin{tablenotes}
      \small
      \item Upper panel: best-fit parameters of the Gaussian model components showed in Figure \ref{fig:spectrum}: central velocity ($\mu_v$), flux amplitude ($S_{peak}$), and velocity dispersion ($\sigma_v$). Lower panel: total CO(1-0) line flux ($S_{CO}$), total CO(1-0) line luminosity ($L'_{CO}$) and molecular gas mass ($M_{H2}$). Although the amplitude of components 2 and 5 is about half of that of the other components, they are significant at a level $\gtrsim 30\,\sigma_{rms}$. The errors on the Gaussian parameters are given at one standard deviation. The uncertainty on the total flux is dominated by the ALMA calibration error of 6\%. The error on the luminosity depends also on the error of the Hubble constant: we used 70 $\pm$ 5 \si{km.s^{-1}.Mpc^{-1}}. The molecular gas mass has the largest uncertainty due to the conversion factor which has a precision of 68\%.
    \end{tablenotes}
    \end{threeparttable}
\end{table}

\begin{figure}
	\includegraphics[width=\columnwidth]{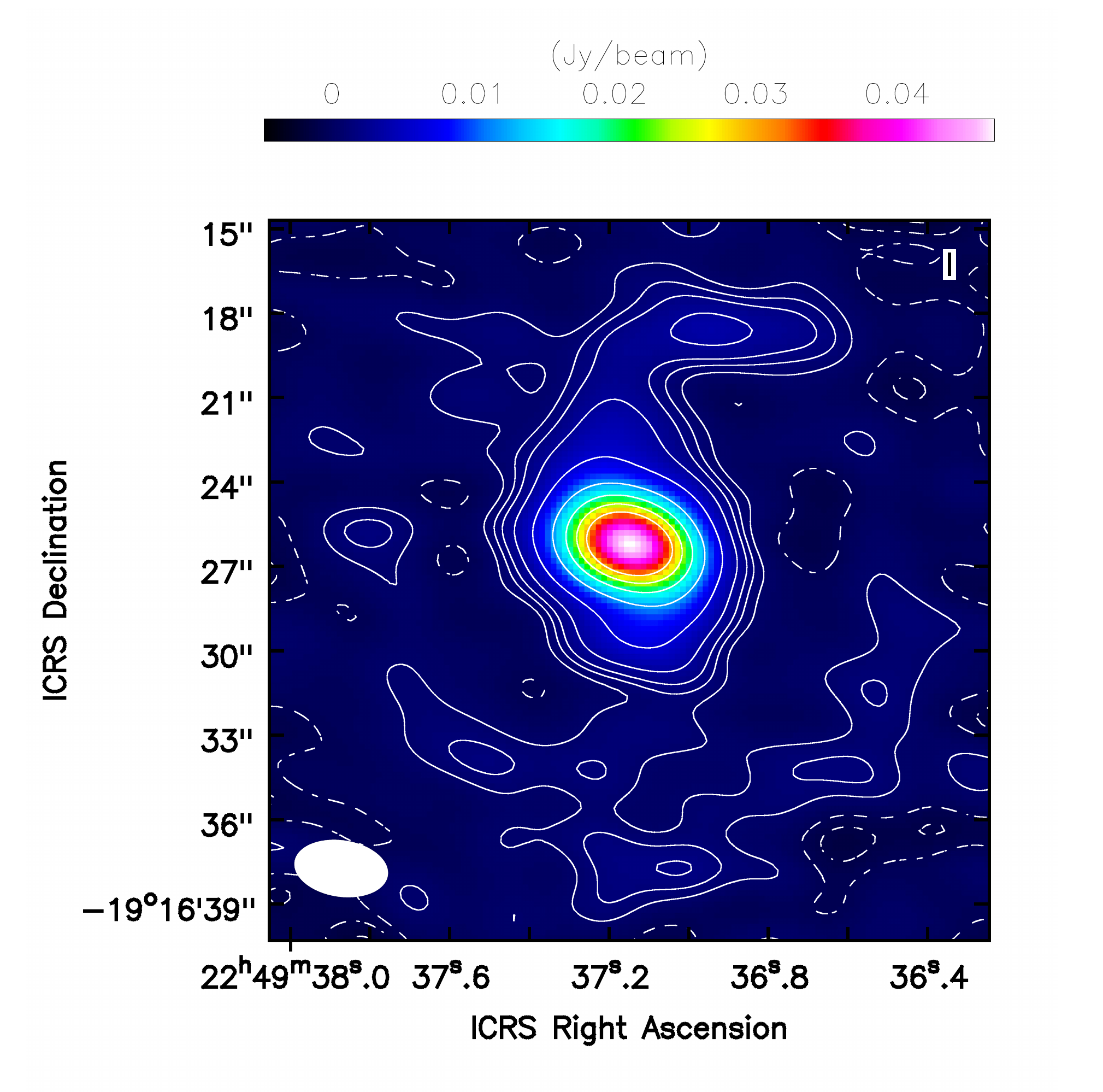}
    \caption{Collapsed map of the CO(1-0) emission: the flux is summed over the spectral range (-238, 213) $\rm km\,s^{-1}$. The white solid lines are the contours level at 1-2-3-5-10-20-30-40-50 $\sigma$, with $\sigma=0.6\,\rm mJy$. Negative contours are also displayed with white dashed lines. The synthesized beam is plotted in the lower left corner in white.}
    \label{fig:collMap}
\end{figure}

In Figure \ref{fig:collMap} we show the collapsed map of the CO(1-0) line emission, obtained by selecting the channels in the range (-238, 213) $\rm km\,s^{-1}$, which excludes the CO(1-0) emission lower than 3$\sigma$. The bulk of the emission is produced in a central circular region of the galaxy of radius 2 kpc  and it is detected with a significance of $\sim50\,\sigma$. Some fainter clumps detected at $2-5\,\sigma$ are located at larger distances, up to 8 kpc away from the centre.

\begin{figure*}
	\includegraphics[width=\linewidth]{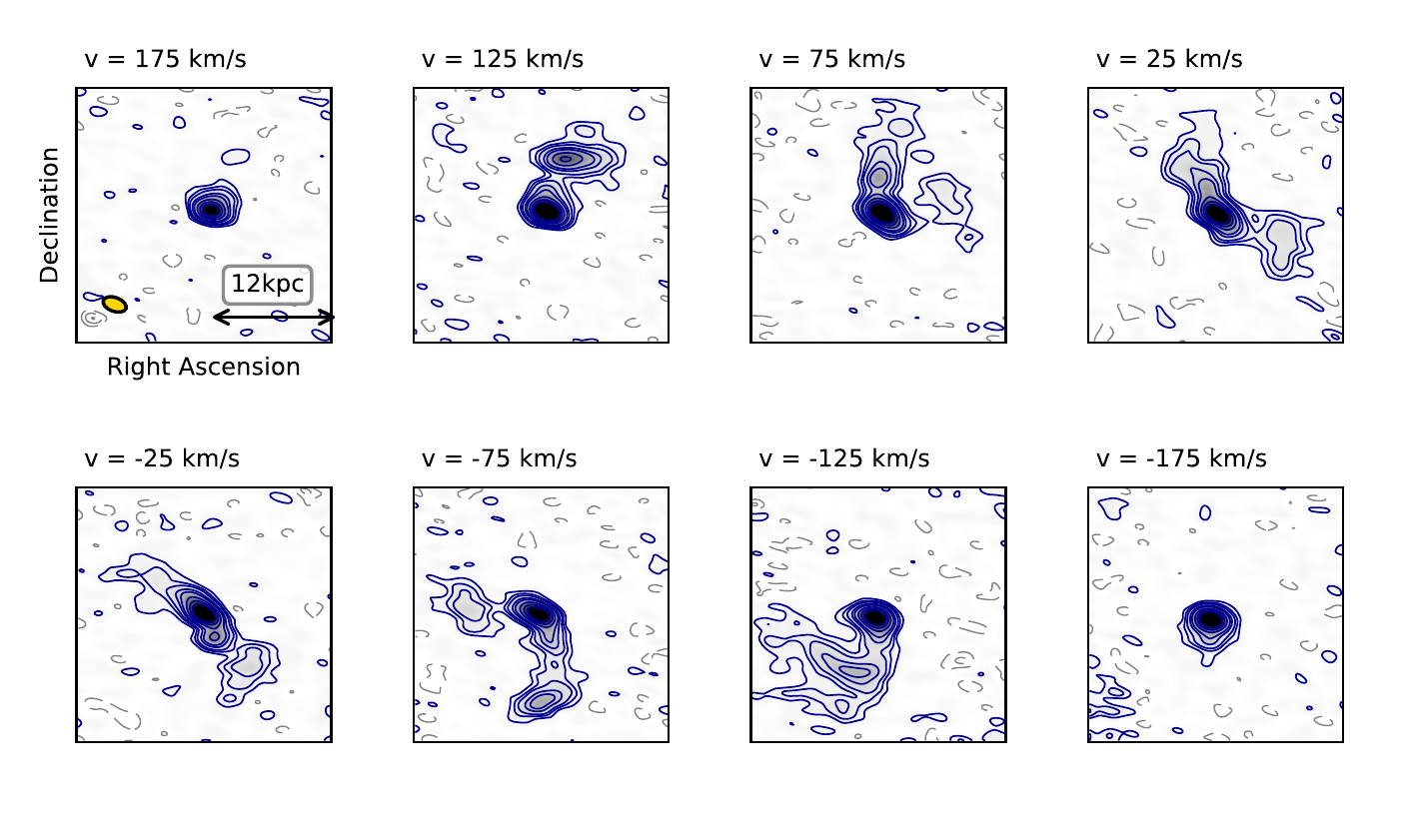}
    \caption{Channel maps of the CO(1-0) emission, with contours at 3-6-9-15-20-30-40-50 $\sigma_{rms}$, $\sigma_{rms}=0.7\,\rm mJy$ (solid blue lines). Negative contours are plotted with grey dashed lines. Channels have a width of 50 $\rm km\,s^{-1}$ and their central velocity is displayed in the upper right corner of each image. The synthesized beam is plotted in yellow in the first panel where we also show the physical scale. Each panel is 35$\arcsec$ x 35$\arcsec$ (23.4 kpc x 23.4 kpc). North and East directions correspond to top and left-hand sides of the panels, as in Figure \ref{fig:collMap}.}
    \label{fig:ChMap}
\end{figure*}

In Figure \ref{fig:ChMap} we show the CO(1-0) channel maps, obtained with $\Delta v = 50\,\rm km\,s^{-1}$. As shown in Figure \ref{fig:ChMap}, the red-shifted emission arises from the half portion of the galaxy oriented N-W, while the blue-shifted emission comes from the other half oriented S-E. This is the typical pattern of an emitting source that is rotating on a disk, inclined with respect to the line of sight.

At the sensitivity of our data, we did not detect any significant CO(1-0) emission at velocities $|v|>250$ $\rm km\,s^{-1}$, i.e. higher than the maximum rotation velocity (see also Figure \ref{fig:spectrum} and \ref{fig:PV}). However, we cannot exclude the presence of a low-velocity outflow whose emission is blended with that of the rotating disk. Therefore in the next Section, we model the emission from the rotating gas, subtract it from the ALMA data-cube and study the morphology and kinematics of any residual CO emission.

\subsection{Modelling of the main rotating disk}\label{sec:disk}





\begin{figure}
	\includegraphics[width=\columnwidth]{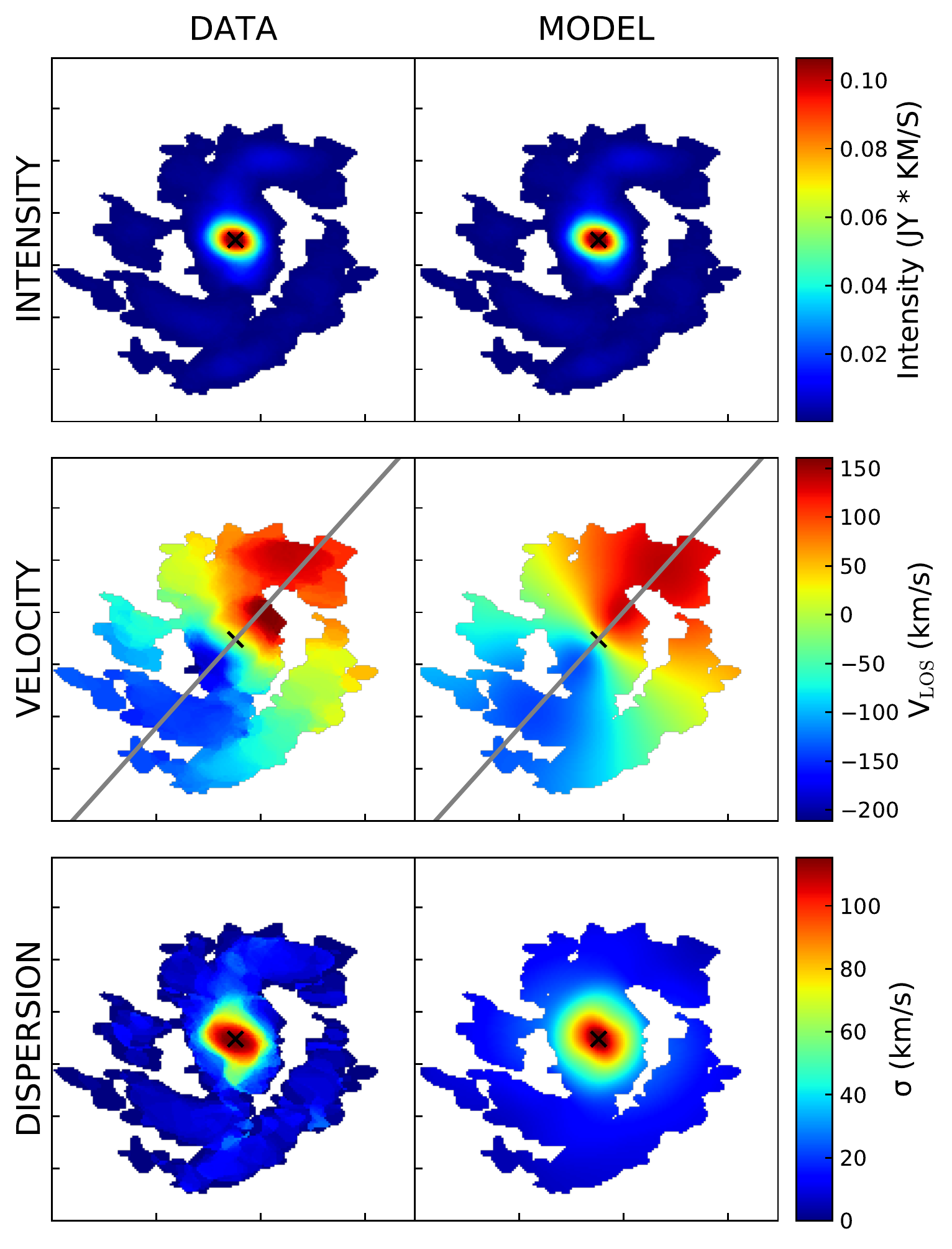}
    \caption{Moment maps of order 0, 1, and 2 (intensity, velocity and velocity dispersion maps, respectively) of the ALMA CO(1-0) data (left panels) and of the best-fit model constructed with \textsc{3D Barolo} (right panels). A black cross marks the centre of the galaxy. The major axis of the molecular disk is shown by a grey solid line in the velocity map. The white pixels are those with an intensity lower than a threshold of $3\sigma_{rms}$.}
    \label{fig:data-model_Maps}
\end{figure}

The CO(1-0) line emission is dominated by the rotation of the gas in the molecular disk, whose axis is inclined with respect to the line of sight. The typical pattern of such emitting disk \citep{Fraternali2002AJ....123.3124F} is clearly visible in the PV diagram of Figure \ref{fig:PV} and in the channel maps in Figure \ref{fig:ChMap}. Before modelling the rotating molecular disk, we inferred its geometrical parameters from the intensity (0th moment) and velocity (1st moment) maps of the CO(1-0) emission, shown in Figure \ref{fig:data-model_Maps} (left panels). The coordinates of the centre of the disk are determined by performing a Gaussian fit of the central part (3 kpc) of the intensity map. The peak emission is located at RA = $22h\,49m\,37.16s\,\pm\,0.07\,s$, Dec = $-19^\circ16'26.4\arcsec\,\pm\,1.0\,\arcsec$, with errors given by the semi-minor axis of the synthesized beam. The inclination angle of the disk has been estimated by measuring the ratio between the minor and major axes of the molecular disk on the velocity map, which features a velocity gradient in the south-west to north-east direction. The inclination angle is defined as the angle between the disk axis and the line of sight (a face-on disk has $i=0^\circ$, an edge-on disk has $i=90^\circ$). We derived a value of $i=37\pm10\,\rm deg$, consistent with the broad range of values inferred using the same method for the optical and IR imaging \citep{jones20096df}. The position angle of the major axis of the disk is also derived from the velocity map: we find a value of $PA=42^\circ$, measured clockwise starting from the north direction.

We used the software 3D-Based Analysis of Rotating Object via Line Observations (\textsc{3D Barolo}), developed by \citet{DiTeodoro2015MNRAS.451.3021D}, for constructing a disk model and fitting its emission to the ALMA data. The main assumption of this model is that all the emitting material of the galaxy is confined to a geometrically thin disk and its kinematics is dominated by pure rotational motion. The typical approach, until recent years, for studying the gas kinematics in a galaxy consisted of fitting the 2D-velocity map with an analytical function for the line of sight velocity. Instead, in the 3D approach, the fitting technique is based on a Monte Carlo extraction of the positions and velocities of the gas clouds. This permits to generate the disk one ring at a time and, after the convolution of the disk model with the observational Point Spread Function (PSF), to perform the minimization on all the maps (channels) comprising the data-cube.

The input data-cube was obtained by cropping the cleaned, continuum-subtracted ALMA CO(1-0) data-cube (size = 35$\arcsec$ x 35$\arcsec$, velocity range = (-264, 258) $\rm km\,s^{-1}$) using the \textsc{CASA} task \texttt{IMSUBIMAGE}. In order to construct the model disk, we used 8 rings of 1.34 kpc width, which is comparable with the angular resolution scale of the ALMA data. With such configuration we covered a region of 21.4 kpc in diameter, which includes the fainter distant clumps shown in Figure \ref{fig:ChMap}. The input and output parameters of \textsc{3D Barolo} are reported in Table \ref{tab:par}.

\begin{table}
    \begin{threeparttable}
    \centering
	\caption{Parameters of the \textsc{3D Barolo} model}
	\label{tab:par}
	\begin{tabular}{lccc} 
		\hline
		\noalign{\vskip 0.5mm}
		\multicolumn{4}{c}{Fixed parameters}\\
		\hline
		i & PA & RA(J2000) & Dec(J2000) \\
		$37^\circ$ & $42^\circ$ & 22h49m37.16s & $-19^\circ16'26.4\arcsec$\\
		\hline
		\noalign{\vskip 0.5mm}
		\multicolumn{4}{c}{Output parameters}\\
		\hline
		\noalign{\vskip 0.5mm}
		Ring & R [kpc] & $v_{\rm los}$ [$\rm km\,s^{-1}$] & $v_{\rm rot}$ [$\rm km\,s^{-1}$] \\
		\hline
		\noalign{\vskip 0.5mm}
		1 & 0.67 & $169^{+8}_{-9}$ & $280^{+14}_{-15}$\\
        2 & 2.01 & $164^{+17}_{-17}$ & $273^{+28}_{-28}$\\
        3 & 3.35 & $134^{+18}_{-12}$ & $223^{+29}_{-20}$\\
        4 & 4.69 & $130^{+6}_{-7}$ & $217^{+10}_{-12}$\\
        5 & 6.03 & $94^{+10}_{-11}$ & $156^{+17}_{-19}$\\
        6 & 7.37 & $134^{+5}_{-6}$ & $222^{+9}_{-10}$\\
        7 & 8.71 & $95^{+7}_{-6}$ & $157^{+12}_{-10}$\\
        8 & 10.05 & $127^{+7}_{-8}$ & $211^{+12}_{-14}$\\
        \noalign{\vskip 0.5mm}
        \hline
        \noalign{\vskip 0.5mm}
        \multicolumn{4}{c}{CO flux and H$_2$ mass of the modelled disk}\\
        \hline 
        \noalign{\vskip 1mm} 
        & $S_{\rm CO}$ & $L^{\prime}_{\rm CO}$ & $\rm M_{H2}$ \\
        & [Jy~km~s$^{-1}$] & [K~km~s$^{-1}$~pc$^{2}$] & [M$_{\odot}$] \\
        \hline
        \noalign{\vskip 0.5mm}
        & $50\pm3$ & $(2.3\pm0.4)\cdot 10^9$ & $(7\pm5)\frac{\alpha_{CO}}{3.1}\cdot10^9$\\
        \hline
	\end{tabular}
	\begin{tablenotes}
      \small
      \item Upper table: values of the input (fixed) parameters of the disk model, (inclination angle $i$, position angle of the major axis $PA$, coordinates of the centre of the disk). Middle table: best-fit rotation parameters for each ring. We report its deprojected radial distance ($R$), the line of sight velocity ($v_{\rm los}$), and the true rotation velocity ($v_{\rm rot}$). Lower table: total CO(1-0) flux ($\rm S_{CO}$), CO(1-0) luminosity ($\rm L^{\prime}_{CO}$) and molecular gas mass of the rotating disk ($\rm M_{H2}$).
    \end{tablenotes}
	\end{threeparttable}
\end{table}
The fitting was performed considering only pixels with a S/N$\gtrsim3$.
In order to estimate the uncertainties on the best-fit values of the rotation velocities, we employed a Monte Carlo method. After the minimization, the algorithm oversamples the neighborhood of the minimum in the parameter space, by slightly changing the values of the fitting parameters. The error on each parameter is taken as the variation that produces an increase of 5$\%$ in the residuals function. The variation of 5\% is the default value chosen by the authors of the software Barolo 3D  \citep{DiTeodoro2015MNRAS.451.3021D}. In other words, all the possible values of a fitting parameter within its uncertainty guarantee an error of <5\% on the minimum residual function, and therefore on the best-fit model.\footnote{As explicitly stated also in \citet{DiT2015PhD}, there is no direct way in the 3D approach to calculate the errors of the fitted parameters. Albeit this “5$\%$ increase” procedure is not optimal and slightly computationally expensive, it returns errors which are in good agreement with those obtained with more standard methods used for the 2D modelling.}

Figure \ref{fig:data-model_Maps} shows the comparison between the data (left-column panels) and the best-fit model (right-column panels) through the moment (0th, 1st and 2nd) maps.

\begin{figure}
	\includegraphics[width=\columnwidth]{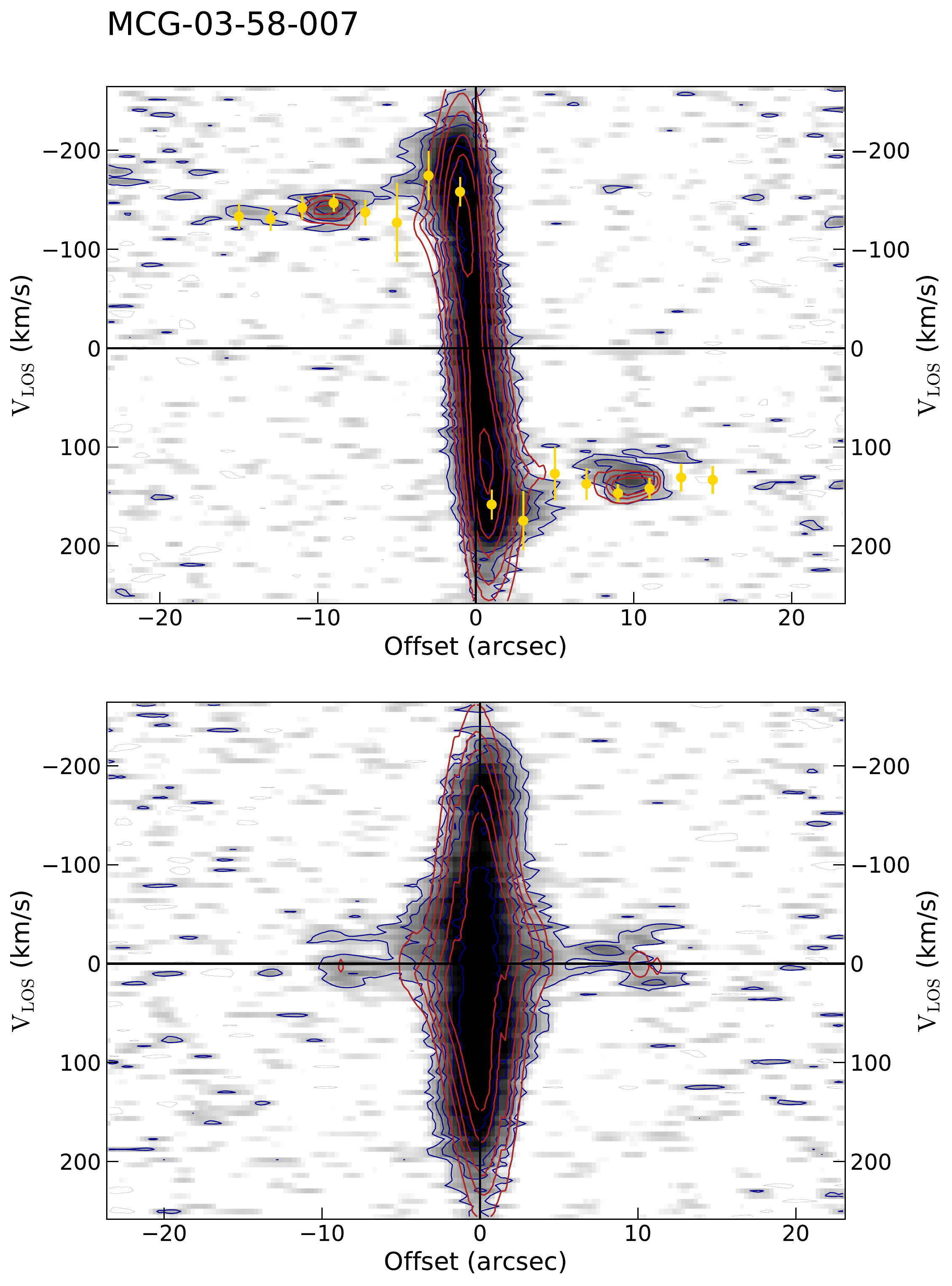}
    \caption{PV diagrams extracted from the data-cube (blue solid contours) and model-cube (red solid contours) using one slit along the major axis (top panel) and another one along the minor axis (bottom panel). The contours level of both the data and the model are at 3-6-9-15-20-30-40-50$\sigma$. The rotation velocity of each ring of the disk model is plotted with yellow solid dots in the top panel along with the error bars.}
    \label{fig:Vrot-PV}
\end{figure}

In Figure \ref{fig:Vrot-PV} we show the PV diagrams extracted from the slits along the major and minor axes of the disk (top and bottom panels respectively). In the top panel we superimpose the LOS rotation velocity of each ring of the model disk. 
 Using the best-fit model, we estimate a total CO(1-0) flux for the rotating molecular disk of $S^{\rm disk}_{\rm CO}=50\pm3\,\rm Jy\,km\,s^{-1}$, from which we derive the CO luminosity of the disk $L^{\prime}_{\rm CO}=(2.3\pm0.4)\cdot 10^9\,$ $\rm{K\,km\,s^{-1}\,pc^{2}}$ and the mass of the molecular disk $M_{\rm H2}=(7\pm5)\frac{\alpha_{CO}}{3.1}\cdot10^9\,M_{\odot}$ (values reported in Table \ref{tab:par}). This corresponds to 78$\%$ of the total molecular gas traced by the CO(1-0) emission detected by ALMA.

\subsection{The CO(1-0) residual emission}

\begin{figure*}
	\includegraphics[width=0.88\linewidth]{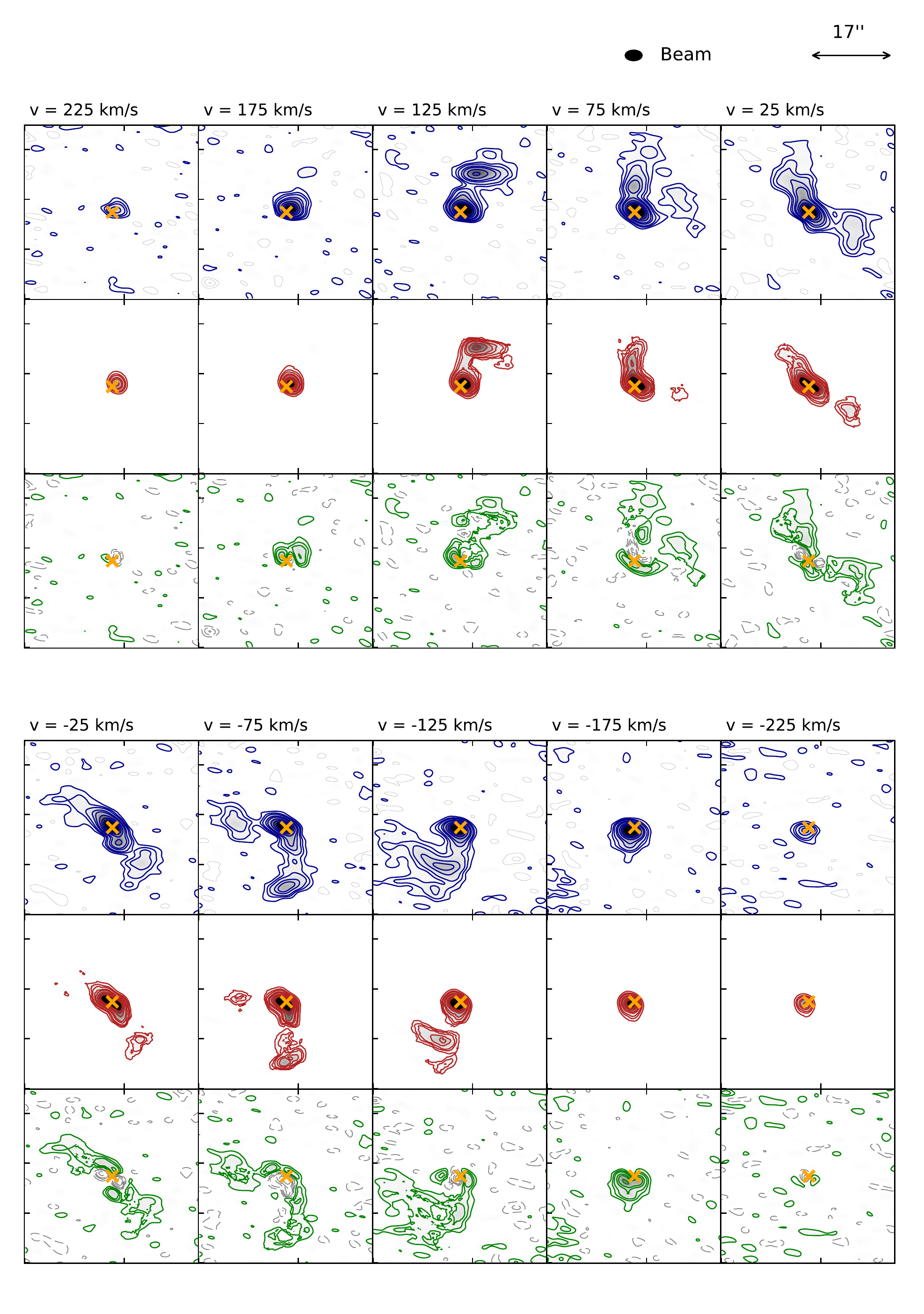}
    \caption{CO(1-0) 50km/s-wide channel maps. From top to bottom rows: data, rotating disk best-fit model, and residuals (blue, red and green solid contours respectively). Contours levels correspond to 3,6,9,15,20,30,40,50~$\sigma_{\rm rms}$ with $\sigma_{\rm rms}=0.7$~mJy. Symmetric negative contours are also displayed with dashed grey lines. The yellow cross in each panel marks the centre of the disk. Each panel is $35\arcsec \times 35\arcsec$ ($23.4\times~23.4$~kpc$^2$). North and East correspond to the upper and left-hand sides of the panels.}
    \label{fig:dat-mod-res}
\end{figure*}

The output of the \textsc{3D Barolo} software is a model-cube of the disk rotation, which we subtracted from the ALMA CO(1-0) line datacube. In Figure \ref{fig:dat-mod-res} we compare data, model, and residuals using channel maps of 50~km~s$^{-1}$ wide channels.

\begin{figure}
	\includegraphics[width=\columnwidth]{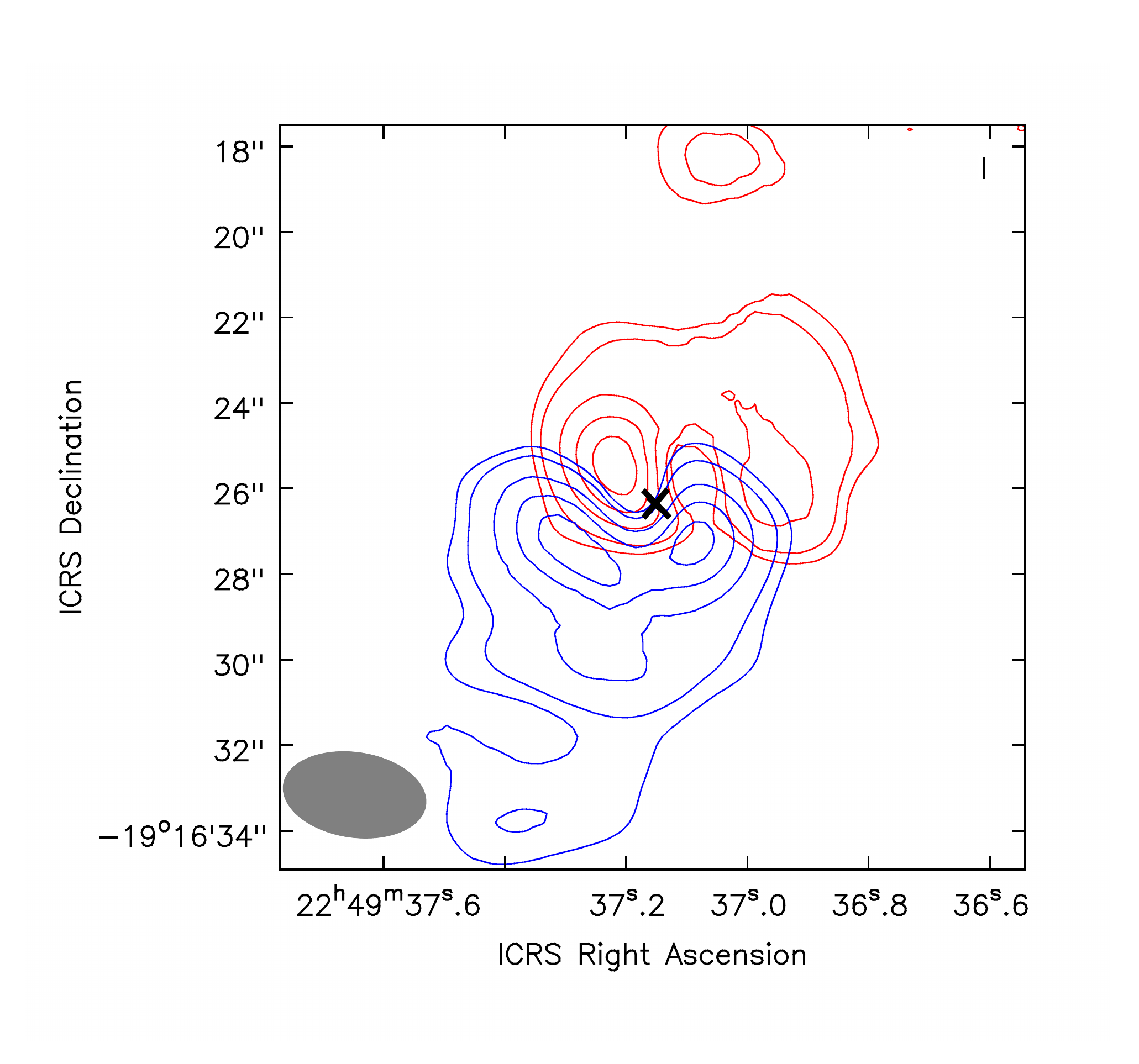}
    \caption{Intensity maps of the blue-shifted (blue solid contours) and red-shifted residual emission (red solid contours) after subtracting the main disk model. The CO emission was integrated respectively over $v\in(-148,-222)$~$\rm km\,s^{-1}$ and $v\in (121, 195)$~$\rm km\,s^{-1}$. Contour levels correspond to 3-5-10-15-20~$\sigma$. The centre of the galaxy is marked with a black cross. The synthesized beam is plotted in grey at the bottom left corner of the panel.}
    \label{fig:RedBlueCont}
\end{figure}

In both the receding and approaching velocities, the residuals show a spatially extended emission detected at the $3\sigma$ level, which is consistent with low-level residual rotation that is not accounted for by the fit. 
In addition, in the two symmetric channels at higher velocities ($-175$~km~s$^{-1}$ and $+175$~km~s$^{-1}$), we detect a compact two-fold structure at much higher significance (S/N=20). In Figure \ref{fig:RedBlueCont} we show a contour map of such compact residual structure, obtained by integrating the residual flux over the spectral ranges $(-222,-148)\,\rm km\,s^{-1}$ and $(121,195)\,\rm km\,s^{-1}$. 

\begin{figure}
	\includegraphics[width=\columnwidth]{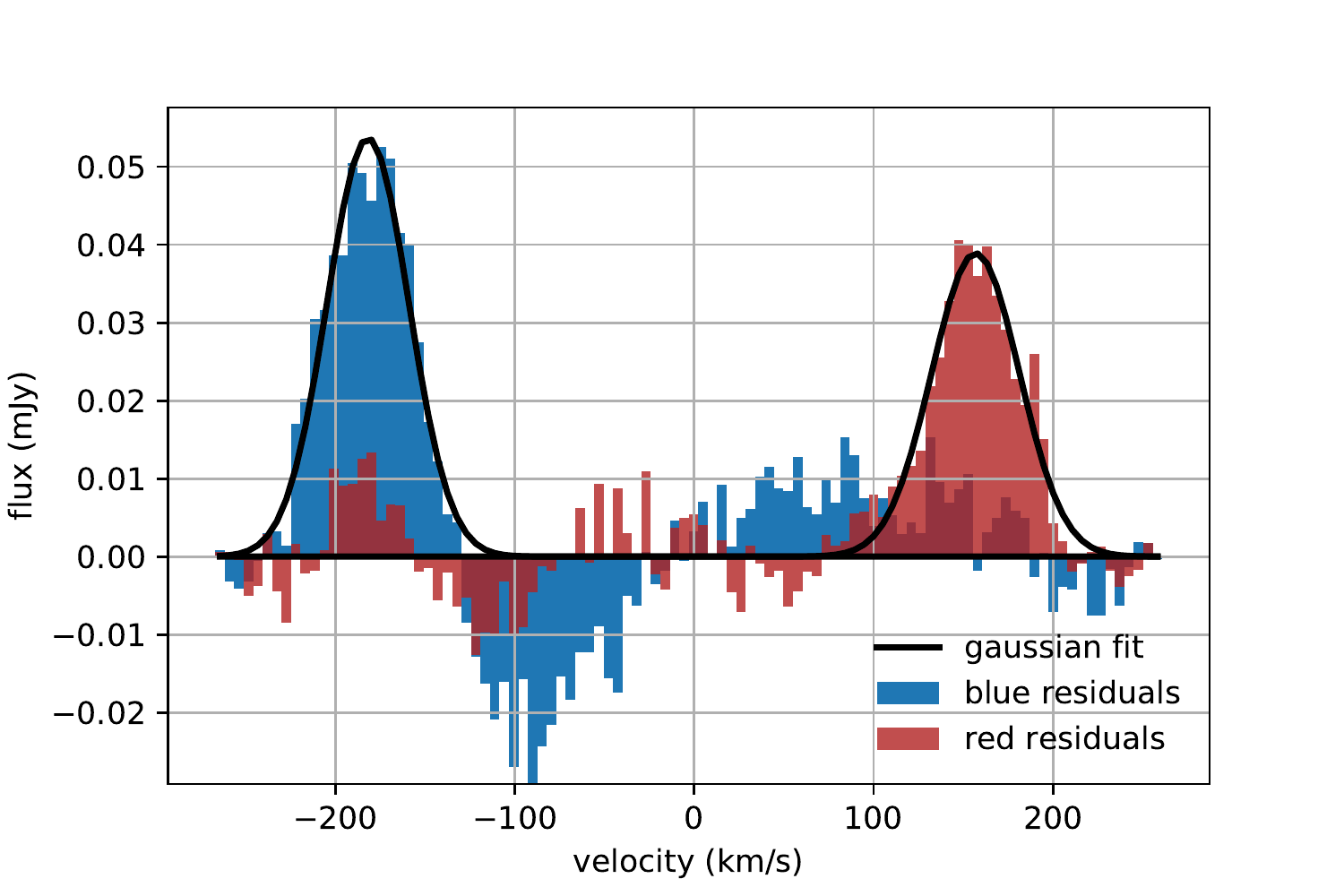}
    \caption{CO spectra extracted from the regions of the blue- and red-shifted emission shown in Figure~\ref{fig:RedBlueCont}. Both spectra were extracted from a polygon area whose perimeter coincides with the contour line at 5$\sigma$ in Figure \ref{fig:RedBlueCont}. The black solid lines show the two best-fit Gaussian functions.}
    \label{fig:residualSpectrum}
\end{figure}

The blue- and red-shifted portions of such structure can fit inside a rectangle of sides 3.3 kpc x 4.7 kpc. The centroids of these two regions are $\sim2.5\arcsec$ (1.7 kpc) offset from the centre of the galaxy. From these two regions we separately extracted the spectra shown in Figure \ref{fig:residualSpectrum}. We modelled both spectra with a Gaussian and computed the fluxes in Table \ref{tab:residuals}, where we report also the respective luminosities and physical sizes. In Section \ref{sec:discussion} we discuss two possible interpretations for such residual compact emission that is not accounted for by the modelling of the main large-scale molecular disk.

\begin{table*}
    \begin{threeparttable}
    \centering
	\caption{Properties of the residual CO(1-0) emission shown in Figure \ref{fig:RedBlueCont} and \ref{fig:residualSpectrum} obtained from the spectral fit.}
	\label{tab:residuals}
	\begin{tabular}{lccccr} 
		\hline
		Gaussian & $v$ ($\rm km\,s^{-1}$) & $S_{CO}$ (Jy $\rm km\,s^{-1}$) & $L'_{CO}$ ($\rm{K\,km\,s^{-1}\,pc^{2}}$) & $R$ (kpc) \\
		\hline
		Red & 157 $\pm$ 5 & 2.39 $\pm$ 0.14 & (1.09 $\pm$ 0.17) $\cdot 10^8$  & 1.7 $\pm$ 0.7\\
		Blue & -182 $\pm$ 5 & 3.07 $\pm$ 0.18 & (1.40 $\pm$ 0.22) $\cdot 10^8$  &  1.7 $\pm$ 0.7\\
		\hline
	\end{tabular}
	\begin{tablenotes}
	\small
	\item  We report the velocity ($v$), the CO flux ($S_{CO}$), the CO luminosity ($L'_{CO}$) and the projected size ($R$) for the two components of the residual emission (red and blue). The projected sizes are defined to be the distance of the centroid to the centre of the galaxy and their errors are given by the semi-minor axis of the synthesized beam. For the uncertainties of other parameters, see caption of Table \ref{tab:tot_flux}.
	\end{tablenotes}
	\end{threeparttable}
\end{table*}

\section{Discussion}\label{sec:discussion}
Our ALMA CO(1-0) observations of MCG-03-58-007 revealed a rotating galaxy-size molecular disk, whose emission contributes to $\sim78\%$ of the total CO(1-0) flux, in addition to a more compact ($R\lesssim2\,\rm kpc$) emission that is not consistent with the modelled rotation. This residual emission, which shows a double-lobe blue-shifted southern component and a symmetrical, double-lobe red-shifted northern component (see corresponding map and spectrum shown in Figures \ref{fig:RedBlueCont} and \ref{fig:residualSpectrum}), may trace a low-velocity outflow ($v\sim170~\rm km~s^{-1}$). Alternatively, this emission may be due to a rotating structure that is decoupled from the main disk fitted by \textsc{3D Barolo}, possibly because it has a  different inclination angle or alignment compared to the large-scale rotation. In the next sections we discuss these two possible interpretations.

\subsection{Case I: a low-velocity compact outflow}\label{sec:h1}

We discuss here the hypothesis that the structure in Fig.~\ref{fig:RedBlueCont} traces a molecular outflow. This interpretation is supported not only by the detection of a powerful X-ray nuclear UFO in this source (as discussed in Sections~\ref{sec:intro} and \ref{sec:target}), but also by the presence of an ionised outflow on scales comparable to the CO residual emission \citep{braito2018new}. In particular, the 6dF optical spectrum \citep{jones20096df} of MCG-03-58-007 shows blue-shifted wings in both the [OIII]4959$\lambda$ and [OIII]5007$\lambda$ emission lines, shifted by about 500~km~s$^{-1}$.
We note that the 6dF optical spectrum was extracted using a fibre aperture of $6.7\arcsec$ (4.5 kpc), which coincides with the region where we detect significant CO residuals (Figure~\ref{fig:RedBlueCont}). However, 
we cannot exclude that the ionised wind extends further up to the kpc scales, outside the fibre coverage. 

In the hypothesis that the residual emission shown in Figure~\ref{fig:RedBlueCont} traces a low-velocity ($v <200$ $\rm km\,s^{-1}$) molecular outflow, we assume a simple biconical model to account for both the blue-shifted and red-shifted emissions that are symmetric with respect to the AGN location (Figure \ref{fig:RedBlueCont}). We derive all the quantities that characterise the outflow separately for the approaching and the receding gas, with the following methods. The outflowing velocities (i.e. $v_{out}^{blue}$ and $v_{out}^{red}$) are estimated from the peaks positions of the best-fit Gaussians of Figure \ref{fig:residualSpectrum}, reported in Table \ref{tab:residuals}. We estimated the size of the outflow components ($R$) as the the projected extensions of the single residual components (red and blue), also reported in Table \ref{tab:residuals}. The dynamical time of the outflow is calculated as $\tau_{dyn} = R/v_{out}$, assuming for simplicity that the axes of the model cones form an angle of 45$^\circ$ with the line of sight. This is equivalent to assuming that the ratio $R/v_{out}$ (projected and line-of-sight component for the extension and the velocity, respectively) represents the ratio between the true extension and velocity of the outflow. Realistic deviations of such angle ($\delta\theta\lesssim 15^\circ$) do not affect the outflow properties beyond their margins of error. 

The masses are derived from the residual CO(1-0) luminosities of Table \ref{tab:residuals} assuming a CO-to-$\rm H_2$ conversion factor of $\alpha^{out}_{CO} = 2.1\,\rm{M_{\odot}(K\,km\,s^{-1}\,pc^{2})^{-1}}$, which is the value calculated by \citet{Cicone2018ApJ...863..143C} for the molecular outflow of NGC 6240\footnote{In NGC 6240 the $\alpha^{out}_{CO}$ estimated for the outflow is lower than the $\alpha_{CO}$ estimated for the non-outflowing $\rm H_2$ gas reservoir, as expected if the large $\sigma_v$ of the clouds in outflow partially decrease the optical depth of its global low-J CO emission \citep{Cicone2018ApJ...863..143C}}. We cannot quote a formal statistical error for the $\alpha^{out}_{CO}$ parameter in MCG-03-58-007's putative outflow (and so all the quantities depending on it). However, based on the few observational constraints available for other molecular outflows in galaxies with similar ISM properties and star formation activity as our target \citep[e.g.][]{Cicone2018ApJ...863..143C,Leroy2015ApJ...814...83L,Zschaechner2018ApJ...867..111Z}, we expect the $\alpha^{out}_{CO}$ parameter to be within $\alpha^{out}_{CO}\in[0.8, 3.4]\,\rm {M_{\odot}(K\,km\,s^{-1}\,pc^{2})^{-1}}$. We note that there is another source, IC5063, for which observations suggest a lower value of $\alpha^{out}_{CO} = 0.3\,\rm {M_{\odot}(K\,km\,s^{-1}\,pc^{2})^{-1}}$ \citep{Oosterloo2017A&A...608A..38O}. However, IC5063 is a Seyfert 2 hosted by an early type galaxy, hence its ISM conditions are very different from our target.

The mass-loss rate in this model is given by $\dot{M}_{out} = M_{out}/\tau_{dyn}$. The momentum rate and the kinetic power of the outflow are defined as $\dot{P}_{out} = \dot{M}_{out}~v_{out}$ and $\dot{E}_{out} = \frac{1}{2}\dot{M}_{out}~v_{out}^2$ respectively. Their estimates are corrected for the line-of-sight effects on the velocity, consistently with the assumption of a bi-cone inclination of 45$^{\circ}$. All the derived quantities of the total molecular outflow are reported in Table \ref{tab:outflows}, along with the same parameters derived for the nuclear wind detected in the X-ray spectrum \citep{braito2018new}. The velocity, radius and dynamical time of the outflowing molecular gas are the values averaged between the approaching and receding components.

\begin{table*}
    \begin{threeparttable}
    \centering
	\caption{Comparison between the molecular outflow and X-ray UFO properties}
	\label{tab:outflows}
	\begin{tabular}{lccccccc} 
		\hline
		Type & $v$ & $R$ & $\tau_{dyn}$ & $M_{out}$ & $\dot{M}_{out}$  & $\dot{P}_{out}$  & $\dot{E}_{out}$ \\
		    &     &  [pc]  & [10$^7$~yr]  & [10$^8$~M$_{\odot}$] & [M$_{\odot}$~yr$^{-1}$] & [g cm s$^{-2}$] & [erg s$^{-1}$] \\
		(1) & (2) & (3) & (4) & (5) & (6) & (7) & (8) \\
		\hline
		\noalign{\vskip 0.5mm}
		$\rm H_2$ outflow \textbf{(*)}& (170 $\pm$ 4)$\,\rm km\,s^{-1}$ & ($1.7\pm0.7$) $\cdot 10^{3}$ & $1.0$  & $5.2\frac{\alpha_{CO}}{2.1}$ & $54\frac{\alpha_{CO}}{2.1}$ & $8\cdot 10^{34}$ & $1.0 \cdot 10^{42}$\\
		X-ray UFO & $0.075c$ & $1.7 \cdot 10^{-3}$ & - & - & $1.5$ & $2 \cdot 10^{35}$ & $2.4 \cdot 10^{44}$\\
		\hline
	\end{tabular}
	\begin{tablenotes}
	\item (1) Outflow type; (2) outflow velocity (LOS component); (3) projected extension of the outflow; (4) dynamic time of the outflow; (5) outflowing mass; (6) mass-loss rate; (7) momentum rate; (8) kinetic power.\\
	(*) The $\rm H_2$ outflow energetics depends on the $\alpha^{out}_{CO}$ conversion factor, which we assumed equal to $2.1\,\rm {M_{\odot}(K\,km\,s^{-1}\,pc^{2})^{-1}}$. As explained in Section \ref{sec:discussion}, based on the few observational constraints available for $\alpha^{out}_{CO}$ in outflow, we expect this value to be comprised within $\alpha^{out}_{CO}\in[0.8, 3.4]\,\rm {M_{\odot}(K\,km\,s^{-1}\,pc^{2})^{-1}}$.
	\end{tablenotes}
	\end{threeparttable}
\end{table*}

\subsubsection{Testing the AGN feedback scenario}

The mass loading factor $\eta\equiv\dot{M}_{out}/SFR\sim3$ indicates that such putative molecular outflow is potentially able to quench the star-formation in MCG-03-58-007. The mass loading factor also suggests that SF alone may not be sufficient at driving such outflow, hence requiring the additional energy from the AGN, as expected from a blast-wave feedback mechanism initiated by the X-ray wind (see also Section \ref{sec:SF}). Therefore, MCG-03-58-007 is a good laboratory to test AGN feedback models, which can be done by comparing the energetics of the putative molecular outflow to the energetics of the stable component of the X-ray UFO. 

The ratio between the momentum rates of the two outflows is $\dot{P}_{mol}/\dot{P}_{UFO}\sim0.4$. Although this ratio has been obtained following a series of assumptions used to estimate the energetics of the two outflows, it allows us to rule out the presence of a momentum-boosted molecular outflow in our target. Furthermore, it is also possible - we note - that a non-negligible fraction of the initial momentum is transferred to the ionised outflow. However, with the current, spatially unresolved, optical data it is not possible to constrain the energetics of the ionised outflow (and, in any case, it is very difficult to constrain the masses and energetics of ionised outflows in absence of an estimate of the electron density, see considerations in e.g. \citet{Harrison2018NatAs...2..198H}). Instead the kinetic power of the molecular outflow is three orders of magnitude lower than that of the X-ray wind. Remarkably, this is at tension with a classic energy-conserving blast-wave feedback scenario which predicts not only the kinetic power to be conserved but also the momentum rate of the large scale outflow to be boosted \citep{Faucher2012MNRAS.425..605F}. The efficiency factor, which we define as the ratio between the kinetic power of the molecular outflow to the X-ray wind, is $\dot{E}_{mol}/\dot{E}_{UFO}\sim4\cdot10^{-3}$, lower than the range found by \citet{Mizumoto2019ApJ...871..156M} for a sample of six Seyfert galaxies, i.e. from $\sim7\cdot10^{-3}$ to $\sim1$. Our results are consistent with the model of a momentum-conserving outflow where the nuclear wind is able to dissipate the energy via a radiative shock \citep{king2010black}. We note that, using the non-relativistic Compton cooling in their hydro-chemical simulations, \citet{Richings2018MNRAS.474.3673R} found that AGN energy-driven outflows can loose the thermalized mechanical energy through efficient cooling and in-situ formation of molecular gas within the outflow, which would imply momentum conservation and energy dissipation similar to what we found for MCG-03-58-007.

Alternatively, the energetics of the molecular outflow might be unrelated to the nuclear X-ray wind. This can be the case if the radiation pressure on the dust of the ISM is the driving mechanism of the large scale outflow \citep{Ishibashi2018MNRAS.476..512I,Costa2018MNRAS.479.2079C}. However, this coupling mechanism is most efficient at $L_{AGN}\gtrsim10^{47}\,\rm erg\,s^{-1}$ whereas the $L_{AGN}$ of MCG-03-58-007 is only $\sim10^{45}\,\rm erg\,s^{-1}$. Furthermore, the presence of a nuclear X-ray UFO favours an interpretation in the framework of blast-wave feedback models.

Another possibility, which we cannot test with the current data, is that this putative molecular outflow is driven by a hidden kpc-scale radio jet (e.g. see \cite{Wagner+12} for a theoretical framework, and \cite{Morganti+15} for observational evidence). A similar case was recently identified by \cite{Husemann2019arXiv190510385H} in HE-1353-1917, where the multiphase outflow, including a CO component, extends by $\sim1$~kpc and has a velocity of $\sim200\,\rm km\,s^{-1}$, similar to MCG-03-58-007.

In any case, our results appear to rule out the hypothesis of a momentum-boosted molecular outflow in MCG-03-58-007, despite the presence of a powerful X-ray wind. We note that a similar result, which is at odds with energy conserving models, has been recently obtained for the most luminous local AGN, PDS~456. In this source, \citet{Bischetti2019arXiv190310528B} detected a large-scale molecular outflow with a momentum rate of the same order of that of the X-ray wind. Furthermore, in IRAS~F11119+3257, new ALMA data allowed to revise the estimate of the momentum rate of the molecular outflow, finding that this source may be more consistent with the momentum-conserving feedback scenario, contrary to previous claims \citep[see][]{Veilleux2017ApJ...843...18V}. Finally, \citet{Mizumoto2019ApJ...871..156M} measured a wide range of values for the ratios between the kinetic powers of molecular outflows to the X-ray UFOs, studying a sample of six ULIRGs and quasar-hosts where both an X-ray wind and a molecular outflow have been detected. These findings, together with our new results, point out that the relation (if any) between the disk-wind and the large-scale ISM is more complex than commonly assumed.

\subsubsection{The role of SF feedback}\label{sec:SF}
When comparing the putative molecular outflow energetics with the AGN luminosity and the nuclear wind energetics, we implicitly assumed that all the feedback processes occurring in MCG-03-58-007 are due to the AGN activity. In order to test this assumption, we need to verify that the kinetic power output due to star formation activity - stellar radiation and SuperNovae (SNe) explosions - is not sufficient to power the putative molecular outflow. The latter extends up to $R_{OF} \sim 2$ kpc (Table \ref{tab:outflows}). Since we have only a measurement of the total SFR throughout the entire extent of the galaxy, we need to scale down the total SFR to take into account that the putative molecular outflow is much less extended than the molecular disk ($R_{gal}\sim9\,\rm kpc$). To do so, we use the $L_{CO}-SFR$ correlation computed for a sample of local SF galaxies \citep{Cicone+17}, which is a direct result of the Schmidt-Kennicutt empirical law \citep{Kennicutt1998ApJ...498..541K}. From the $L_{CO}$ measured in the nuclear region at $r<(2\pm1)\,\rm kpc$, we infer SFR$(r<2\,\rm kpc) = 10 \pm 5\,\rm M_{\odot}\,yr^{-1}$. Using the theoretical predictions of the stellar winds and SNe output in the ISM \citep{Veilleux2005ARA&A..43..769V}, which combine stellar evolutionary models, atmospheric models and empirical spectral data, we find that SF activity can trigger a mass-loss rate of $\dot{M}_{SF}=2.7\,\rm M_{\odot}\,yr^{-1}$ within a radius of 2 kpc. This is a factor of 20 lower than what we measured for the putative outflow. The expected kinetic power injected into the ISM by SNe is $\dot{E}_{SN}=7.3\cdot10^{42}\,\rm erg\,s^{-1}$, which can explain the kinetic energy of the molecular outflow only by assuming a coupling efficiency of $14\%$. Such efficiency value is at the high end of the range allowed by theoretical models. Hence, the energetics of the putative outflow would be hard to reconcile with a purely star formation-driven feedback process. 

\subsection{Case II: a kpc-scale rotating structure}\label{sec:h2}

An alternative interpretation for the CO(1-0) residuals shown in Fig.~\ref{fig:RedBlueCont} would be the presence of a compact structure that is rotating possibly with a different geometry and kinematics with respect to the rotation of the larger molecular disk modelled with \textsc{3D Barolo}. The compact CO(1-0) emission shown in Figure \ref{fig:RedBlueCont} resembles the shape of a doughnut, which may trace a ring with a radius of $R\simeq1.7$~kpc, rotating with LOS velocities of about $v_{\rm los}\simeq170$~km~s$^{-1}$ and having a molecular mass of $M^{ring}_{H2} \simeq 8\frac{\alpha_{CO}}{3.1}\cdot10^8\,\rm M_{\odot}$. However, it is difficult to infer the inclination and position angle of such putative ring, for the following reasons: 

\begin{description}
    \item[(i)] The limited spatial resolution of the ALMA data does not allow us to constrain the geometry of this structure;
    \item[(ii)] The CO residual emission in Fig.~\ref{fig:RedBlueCont}, especially the components that are aligned along the major axis of the main molecular disk fitted by 3D Barolo (NW-SE direction), might still include a contamination from such rotation; 
    \item[(iii)] The kinematics of the CO residuals (Figure~\ref{fig:residualSpectrum}) is likely more complex than a pure rotation pattern and we can provide only a zeroth order description.
\end{description}

Compact rotating structures have been detected at smaller scales ($\sim 10-100\,\rm pc$) in a few Seyfert 2 galaxies and are often denominated circumnuclear disks (CNDs). A closed asymmetrical elliptical ring has been detected in NGC~1068 using maps of CO(3-2) line emission \citep{Garc2014A&A...567A.125G} and $\rm C_2H$ \citep{Garc2017A&A...608A..56G}; this CND has a radius of $r \sim 200\,\rm pc$ and is off-centered relative to the location of the AGN. \citet{Izumi2018ApJ...867...48I} found a $74\,x\,34\,\rm pc$ CND with a molecular mass of $M_{H2}\sim 3\cdot10^6\,\rm M_{\odot}$ in Circinus, the nearest type 2 Seyfert galaxy. \citet{Combes2019A&A...623A..79C} have recently detected six CNDs in a sample of seven local galaxies, with radii ranging from 6 to 27 pc and masses from $0.7\cdot10^7$ to $3.9\cdot10^7\,\rm M_{\odot}$. All six structures show an orientation along the line of sight that is unaligned to that of the host galaxy and, except in one case, they are off-centred with respect to the AGN \citep{Combes2019A&A...623A..79C}. We note however that the structure detected in MCG-03-58-007 is a factor of $\sim 10-100$ larger in size and $\rm H_2$ mass, compared to typical CNDs detected so far.
Based on the comparison with previously detected CNDs, we favour the outflow interpretation for the compact residual CO emission in MCG-03-58-007, but a rotating structure cannot be ruled out based on the current ALMA data.

\section{Summary and conclusions}\label{sec:conclusions}


In this work we reported the ALMA CO(1-0) observations of MCG-03-58-007 ($z_{CO}=0.03236\pm0.00002$, this work), a Seyfert~2 galaxy that we selected as a suitable candidate for investigating AGN feedback mechanisms because of the presence of an extremely powerful X-ray disk-wind \citep{braito2018new,Matzeu2019MNRAS.483.2836M}. The target galaxy has a moderate SFR of $20.1\pm0.9$~$\rm M_{\odot}~yr^{-1}$, which makes it one of the few sources in this more typical SF regime where the blast-wave AGN feedback scenario has been tested, and so where the possible contamination from SF feedback processes is expected to be minimised.

The ALMA data that we presented probe the distribution and kinematics of the molecular ISM in the whole galaxy up to scales of $\sim 20\,\rm kpc$. At the sensitivity of our data, we do not detect any signature of very high-velocity molecular outflows ($v>250$~km~s$^{-1}$). The CO(1-0) emission is dominated by rotation in a molecular disk, which we modelled using the software \textsc{3D Barolo}, measuring a molecular mass of $M^{disk} = (7\pm5)\cdot10^9\,\rm M_{\odot}$, corresponding to 78~\% of the total molecular gas mass in this galaxy.

However, the model does not properly describe the complex CO(1-0) kinematics in the central region of MCG-03-58-007. The data-to-model residuals show a compact structure (within 2 kpc from the galaxy centre) with a $20\sigma$ significance, whose emission shows two symmetric peaks at $v\sim-170\,\rm km/s$ and $v\sim+170\,\rm km/s$. This could be the signature of a low velocity molecular outflow. Under this hypothesis, we applied a simple biconical model and estimated the LOS velocity ($v=170\pm4\,\rm km/s$), mass-loss rate ($\dot{M}_{out}\sim54\,\rm M_{\odot}\,yr^{-1}$), momentum rate ($\dot{P}_{out} \sim 8 \cdot10^{34}\,\rm g\,cm\,s^{-2}$), and kinetic power ($\dot{E}_{out} \sim 1.0 \cdot10^{42}\,\rm erg\,s^{-1}$) of such putative molecular outflow. By comparing these values with the energetics of the X-ray disk-wind, we conclude that the energy coupling between the nuclear wind and the large-scale ISM is not as efficient as it is predicted for the more extreme outflows detected in some ULIRGs \citep{Cicone2014A&A...562A..21C,Fiore2017A&A...601A.143F,Fleutsch2019MNRAS.483.4586F}. In the theoretical framework of the blast-wave AGN feedback scenario \citep{king2010black,Faucher2012MNRAS.425..605F}, our results would be more consistent with a momentum-driven outflow rather than an energy-driven outflow. On the other hand, models invoking the effect of AGN radiation pressure on dusty clouds may provide an alternative explanation for the modest energetics of the putative molecular outflow \citep{Ishibashi2018MNRAS.476..512I,Costa2018MNRAS.479.2079C}, although they would not explain the simultaneous presence of a powerful X-ray UFO in this source. With an estimated SFR$(r<2\,\rm kpc) = 10 \pm 5\,\rm M_{\odot}\,yr^{-1}$,  we cannot rule out a significant contribution of SF activity to the driving mechanism. However, in the hypothesis of a purely SF-driven outflow, the required coupling efficiency would be very high, i.e. $7\%$. Moreover, star formation feedback cannot be responsible for the nuclear X-ray UFO.

The low-velocity molecular outflow hypothesis is not the only interpretation for the compact CO residual emission detected in MCG-03-58-007. Indeed, the current ALMA data do not allow us to rule out that such structure traces a rotating ring or disk whose kinematics is decoupled from the main molecular disk (and so not accounted for by the \textsc{3D Barolo} modelling). However, the size and mass of this putative rotating structure in MCG-03-58-007 would exceed by a factor of $10-100$ the typical values measured for CNDs, which have been detected in other Seyfert galaxies \citep{Garc2014A&A...567A.125G,Izumi2018ApJ...867...48I,Combes2019A&A...623A..79C}. Moreover, in this CND hypothesis, MCG-03-58-007 would show neither a high-velocity nor a low-velocity molecular outflow, which would be even more bizarre given the presence of a powerful X-ray wind and of a kpc-scale ionised outflow \citep{braito2018new}.

Our results, together with other recent works \citep{Bischetti2019arXiv190310528B,Fleutsch2019MNRAS.483.4586F,Cicone2018ApJ...863..143C,Veilleux2017ApJ...843...18V} show that the interpretation of molecular outflows in the framework of AGN feedback models is becoming more and more complicated as we increase sample sizes and study different types of galaxies. However, the targets studied so far do not constitute a complete and unbiased sample, which is necessary to investigate whether massive outflows are regulated by a universal mechanism and, if so, to construct a physical model that describes such mechanism.

\section*{Acknowledgements}
This project has received funding from the European Union's Horizon 2020 research and innovation programme under the Marie Skłodowska Curie grant agreement No.664931. This paper makes use of the following ALMA data: ADS/JAO.ALMA\#2016.1.00694.S. ALMA is a partnership of ESO (representing its member states), NSF (USA) and NINS (Japan), together with NRC (Canada), MOST and ASIAA (Taiwan), and KASI (Republic of Korea), in cooperation with the Republic of Chile. The Joint ALMA Observatory is operated by ESO, AUI/NRAO and NAOJ.
We thank Paola Andreani, Roberto Maiolino and Fabio Castagna for the fruitful discussions. GAM is supported by the European Space Agency (ESA) Research Fellowships.
We thank the referee for providing insightful comments that helped improve the paper.





\bibliographystyle{mnras}
\bibliography{literature} 








\bsp	
\label{lastpage}
\end{document}